\documentclass[longauth]{aa}  

\usepackage{graphicx}
\usepackage{txfonts}
\usepackage{siunitx}
\usepackage{hyperref}
\usepackage{stackengine}
\usepackage{placeins}
\usepackage{float}
                                
\begin{document}

   \title{Investigating the magnetic field in the inter-cluster filament between Abell 3667 and Abell 3651 with POSSUM}
   \titlerunning{Magnetic field in the A3667/3651 inter-cluster filament}
   \authorrunning{C. Stuardi et al.}

   \author{C. Stuardi\inst{1}, S.~P. O'Sullivan\inst{2}, L. Rudnick\inst{3}, G. Bernardi\inst{1,4,5}, A. Bonafede\inst{6,1}, J. Dietl\inst{7}, T. Akahori\inst{8,9}, D. Alonso-L{\'o}pez\inst{2}, C. Anderson\inst{10}, E. Carretti\inst{1}, B.~M. Gaensler\inst{11,12,13}, G. Heald\inst{14}, F. Loi\inst{15}, Y.~K. Ma\inst{16}, E. Osinga, G. Pignataro, C. Riseley\inst{17}, X. Sun\inst{18}, A. Thomson\inst{19,14}, C.~L. Van Eck\inst{10}, T. Vernstrom\inst{14,20} \and J.~L. West\inst{21,22}}

   \institute{INAF - Istituto di Radioastronomia (IRA), Via Gobetti 101, 40129 Bologna, Italy\\
             \email{ccstuardi@gmail.com}
             \and Departamento de F\'isica de la Tierra y Astrof\'isica \& IPARCOS-UCM, Universidad Complutense de Madrid, 28040 Madrid, Spain
             \and Minnesota Institute for Astrophysics, University of Minnesota, 116 Church St SE, Minneapolis, MN 55455, USA
             \and Centre for Radio Astronomy Techniques and Technologies (RATT), Department of Physics and Electronics, Rhodes University, Makhanda 6140, South
             \and South African Radio Astronomy Observatory, Cape Town 7700, South Africa
             \and Dipartimento di Fisica e Astronomia, Universit\`a di Bologna, via P. Gobetti 93/2, 40129 Bologna, Italy
   \and Argelander-Institut f\"ur Astronomie (AIfA), Universit\"at Bonn, Auf dem H\"ugel 71, 53121 Bonn, Germany
   \and Mizusawa VLBI Observatory, National Astronomical Observatory of Japan, 2-21-1 Mitaka, Tokyo 181-8588, Japan
   \and SOKENDAI, Shonan Village, Hayama-machi, Miura-gun, Kanagawa 240-0193 Japan
    \and Research School of Astronomy \& Astrophysics, The Australian National University, Canberra, ACT 2611, Australia 
   \and Department of Astronomy and Astrophysics, University of California Santa Cruz, Santa Cruz, CA 95064, USA
   \and Dunlap Institute for Astronomy \& Astrophysics, University of Toronto, Toronto, ON M5S 3H4, Canada
   \and David A. Dunlap Department of Astronomy \& Astrophysics, University of Toronto, Toronto, ON M5S 3H4, Canada
   \and Australia Telescope National Facility, CSIRO Space \& Astronomy, PO Box 1130, Bentley, WA 6102, Australia
    \and INAF - Osservatorio Astronomico di Cagliari, Via della Scienza 5, 09047 Selargius, Italy
   \and Max-Planck-Institut für Radioastronomie, 
Auf dem Hügel 69, 53121 Bonn, Germany
\and Astronomisches Institut der Ruhr-Universität Bochum (AIRUB), Universitätsstraße 150, 44801, Bochum, Germany
\and School of Physics and Astronomy, Yunnan University, Kunming, 650500, PR China
\and SKA Observatory, SKA-Low Science Operations Centre, 26 Dick Perry Avenue, Kensington WA 6151, Australia
    \and International Centre for Radio Astronomy Research (ICRAR), University of Western Australia, 35 Stirling Hwy, Crawley, WA 6009, Australia
    \and School of Natural Sciences, University of Tasmania, PO Box 807, Sandy Bay, TAS 7006, Australia
    \and Dominion Radio Astrophysical Observatory, Herzberg Astronomy \& Astrophysics, National Research Council Canada, P.O. Box 248, Penticton, BC V2A 6J9, Canada \\ }

\date{Received XX}
 
  \abstract
   {Magnetic fields are expected to permeate the filaments of the cosmic web, but direct observational constraints in individual systems remain extremely limited. Faraday rotation studies based on dense grids of polarised background sources provide one of the most promising probes of these weak magnetic fields.}
   {The objective of this study is to measure the magnetic field within the prominent inter-cluster filament recently detected in X-rays by the extended ROentgen Survey with an Imaging Telescope Array (eROSITA). This filament spans over 13 Mpc projected on the sky, connecting the galaxy clusters \object{Abell 3667} and \object{Abell 3651} at $z\sim0.058$}
   {We employed the Polarisation Sky Survey of the Universe's Magnetism (POSSUM) rotation measure (RM) grid to isolate the RM dispersion and median value of background polarised sources induced by the filament’s magnetised plasma and infer the strength of this magnetic field. The filament region was sampled by 54 background polarised sources.}
   {Following subtraction of the foreground Galactic RM, we detect a marginal residual RM dispersion in the filament region of $\sigma_{\rm RRM,excess}=6.9 \pm 3.6~\mathrm{rad~m^{-2}}$, together with a coherent residual RM signal with median $\widetilde{\rm RRM}_{\rm excess}=6.3 \pm 1.3~\mathrm{rad~m^{-2}}$. Assuming simplified single-scale magnetic-field models and adopting informed priors on the thermal electron density distribution derived from the X-ray analysis, we constrain the magnetic field strength to the range 0.1-3.5 $\mu$G within a 95\% confidence level, with preferred values around $0.2$-$0.3~\mu$G depending on the assumed magnetic-field coherence scale. However, we also find that the Galactic foreground RM in this region is highly structured on angular scales comparable to the extent of the filament itself, representing a major source of uncertainty for the RM analysis.}
   {Our results provide the first magnetic field constraints based on Faraday rotation measurements in an individual X-ray-detected inter-cluster filament, with field strengths consistent with theoretical expectations for gas in bridges and cluster outskirts. Our analysis also highlights the critical importance of accurately modelling Galactic RM foregrounds for future studies of extragalactic magnetism with POSSUM and the Square Kilometre Array.}


   \maketitle
   \nolinenumbers
\section{Introduction}

At the largest scales, the Universe forms a network of filaments and bridges through which matter flows, channelling along gravitational potential wells into galaxy clusters - the nodes of this cosmic web \citep{Bond96}. Optical observations have long identified these filamentary structures of galaxies connecting clusters \citep[see][for some compilations]{Tempel14,Malavasi20b}. More recently, X-ray and Sunyaev-Zeldovich observations uncovered the tip of the iceberg of filament's thermal gas reservoir, detecting filaments in the outskirts of massive galaxy clusters \citep{Eckert15Nat, Veronica24,Breuer25}, between close cluster pairs \citep{Planck13filaments,Reiprich13,Hincks22,Dietl24,Migkas25}, and within stacked filaments \citep{Tanimura19,DeGraaff19,Singari20,Zhang24,Isopi25}. However, the magnetic field properties of this vast and rarefied plasma component remain largely unexplored. 

Revealing signatures of magnetic fields in cosmic web filaments is crucial for understanding the origin of cosmic magnetism, whether primordial \citep{Kulsrud08} or astrophysical \citep{Furlanetto01}, since the two models diverge in their predictions for the strength and distribution of magnetic fields in filaments \citep[e.g.][]{Vazza21}. Actual stacking experiments point to the presence of a primordial magnetic field reaching an average strength of 10-60 nG in local filaments \citep{Carretti25}. If this is the case, rare single-object detection at the level of $\sim100$ nG is expected, and it would probe whether dynamo or shock amplification of primordial seeds is already active in filaments, potentially dominating the signal in stacking experiments.

On the observational side, obtaining constraints on large-scale magnetic fields is a big challenge for modern astrophysics. So far, most studies have been limited to galaxy clusters. Radio observations of diffuse radio sources have demonstrated that the intra-cluster medium is permeated by cosmic rays and magnetic fields, with strengths ranging from 0.1 to 10 $\mu$G \citep{vanWeeren19}. Here, magnetic fields radially decrease with the thermal gas density and exhibit turbulent fluctuations over tens to hundreds of kiloparsecs, statistically described with a power spectrum \citep{Bonafede10,Stuardi21,Osinga22,Osinga25}. These results were obtained through measurements of the Faraday effect.

The Faraday effect is the rotation of the polarisation angle of linearly polarised emission caused by a foreground magneto-ionised medium. The amount of rotation is proportional to the squared wavelength ($\lambda^2$) and the rotation measure (RM). The observed RM depends on the magnetic field properties and thermal electron density of the medium crossed by the radiation:
\begin{equation}
   \text{RM} = 812\int_{{z_{\rm source}}}^{\rm{0}} {\frac{n_e(z) B_{\parallel}(z)}{(1+z)^2} \frac{\text{d}l}{\text{d}z} \text{d}z } \quad {\rm rad \ m^{-2}} ~,
\label{eq:RM}
\end{equation}
\noindent
where $z_{\rm source}$ is the redshift of the source, $n_e$ is the thermal electron number density per cubic centimetre, $B_{\parallel}$ is the magnetic field component parallel to the line of sight in microgauss, and d$l$ is the infinitesimal path length in kiloparsecs. The redshift correction is introduced to transform the observed quantity into the reference frame of the Faraday rotating medium. Conventionally, $B_{\parallel}$ is positive when pointing towards the observer and negative otherwise. 

With a large number of RMs from background sources forming an RM grid, it is possible to detect magnetic fields and density distributions across cosmic structures. However, only a few single objects can be studied with this method, and, more often, the RM signal is detected on a statistical basis. The statistical detection of magnetic fields in filaments was recently claimed by \citet{Carretti22}, who studied the evolution of the RM dispersion with redshift and compared it with the prediction from cosmological simulations \citep{Carretti23, Carretti25}. \citet{Anderson24} detected magnetised gas in the outskirts of a large statistical sample of galaxy groups. \citet{Pignataro25} focused instead on filaments in superclusters and constrained magnetic fields by stacking the RM grid of three local superclusters.

To date, no RM‑grid analysis has been performed on an individual filament for two main reasons. First, constraining magnetic fields via the RM‑grid technique requires prior knowledge of the thermal gas content. Reliable density estimates for filaments and bridges have only become available in the last few years, mainly thanks to the Spectrum Roentgen Gamma (SRG) eROSITA X-ray telescope \citep{Predehl21}, which demonstrates superior response to extended soft X-ray emission. Second, only with the advent of modern radio interferometers have RM grids dense enough for such studies become available. In particular, a survey well-suited for this purpose is the Polarisation Sky Survey of the Universe’s Magnetism \citep[POSSUM,][]{Gaensler25}, conducted with the Australian Square Kilometre Array Pathfinder \citep[ASKAP,][]{Hotan21}, alongside its total intensity counterpart, the Evolutionary Map of the Universe \citep[EMU,][]{Norris11,Hopkins25}. The full POSSUM survey will release a catalogue of polarised sources with a density of 30-50 RM/deg$^2$ over $\sim$20000 deg$^2$ in the southern sky, with a broad frequency coverage (800-1088 MHz and 1296-1440 MHz). Rotation measure grid studies performed with POSSUM early data products have demonstrated the survey's unprecedented sensitivity in the outskirts of galaxy clusters and groups \citep{Anderson21,Anderson24,Khadir26,AlonsoLopez26}. Very recently, a deep MeerKAT \citep{Jonas16} observation of the merging system Abell 3391-3395 was used to derive magnetic field properties in the X-ray bridge connecting the two clusters \citep{Gustafsson26}.

The aim of this work is to perform the first ever RM grid study of a single X-ray detected filament, using POSSUM. The chosen target is the long filament between the nearby galaxy clusters Abell 3667 (A3667, $z=0.0556$) and Abell 3651 (A3651, $z=0.0600$). This paper is structured as follows: we introduce our target in Sect.~\ref{sec:data}. We describe POSSUM data and the methods of our analysis in Sect.~\ref{sec:methods}. We present our results in Sect.~\ref{sec:results} and discuss them in Sect.~\ref{sec:discussion}. Finally, we summarise our work in Sect.~\ref{sec:conclusions}. The assumed cosmology in this work is a flat lambda cold dark matter ($\Lambda$CDM) model, with cosmological parameters $h = 0.7$, $\Omega_m = 0.3$, and $\Omega_\Lambda = 0.7$ \citep{Planck18}. We refer to the average redshift between the two clusters ($z=0.058$) for the physical-to-angular scale conversion (1.123 kpc/$\arcsec$), while the precise conversion for the two clusters is given in Table~\ref{tab:clusters}.

\section{The Abell 3667/3651 system}
\label{sec:data}

\begin{table*}[]
    \centering
    \caption{Clusters properties from \citet{Piffaretti11}.}
    \begin{tabular}{cccccccc}
       Name & R.A.(J2000) & Dec(J2000) & $z$ & $M_{500}$ & $R_{500}$ & $R_{200}$ & scale \\
            &  h:m:s    &   d:m:s    &     &  $10^{14}~M_\odot$ & Mpc & Mpc & kpc/$\arcsec$ \\
       \hline

       Abell 3667  & 20:12:30.5 & -56:49:55 & 0.0556 & 5.17 & 1.199 & 1.845 & 1.080 \\
       Abell 3651  & 19:52:16.5 & -55:03:42 & 0.0600 & 1.47 & 0.788 & 1.213 & 1.159 \\
    \end{tabular}
     \tablefoot{Column 1: Cluster name. Columns 2 and 3: Celestial coordinates. Column 4: redshift. Column 5: $M_{500}$, i.e. the mass contained in a spherical volume whose mean density is 500 times the critical density at the cluster’s redshift. Column 6: $R_{500}$, i.e. the radius of this spherical volume. Column 7: $R_{200}$ has a similar definition for an average over-density of 200, computed following \citet{Reiprich13}. Column 8: Physical-to-angular scale conversion at the cluster's redshift.}
    \label{tab:clusters}
\end{table*}

   \begin{figure*}
        \sidecaption
        \includegraphics[width=12cm]{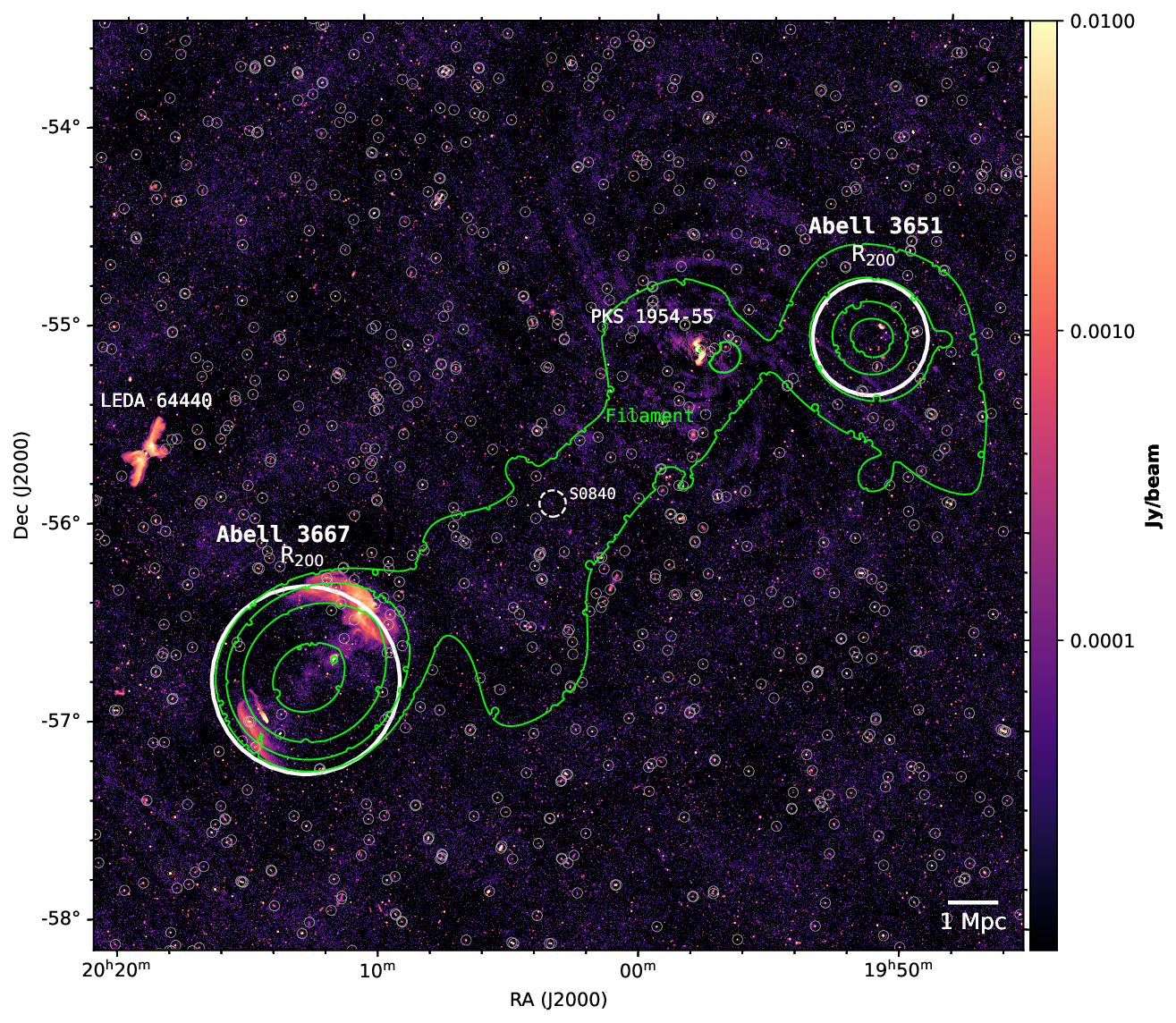}
        \caption{EMU total intensity image centred on the Abell 3667/3651 system at 943 MHz. The white circles mark the R$_{200}$ boundaries of the two Abell clusters, while the dashed white circle represents the region of $4\arcmin$ radius around the S0480 foreground system, excluded from X-ray analysis by \citet{Dietl24}. The names of the two extended galaxies in the field are also reported. The grey circles mark the position of the polarised sources detected by POSSUM. The green contours enclose the region of the filament, as detected by eROSITA \citep{Dietl24}.}
        \label{fig:EMU}
    \end{figure*}

Recently, by using the data from eROSITA All Sky Survey \citep[eRASS1,][]{Merloni24}, \citet{Dietl24} detected soft X-ray thermal emission from the filament between A3667 and A3651. The projected length of this filament is 13 Mpc. However, through the redshift
analysis of the sources between them, which confirms the presence of the filament, they estimate its physical length to be between 25 and 32 Mpc with an inclination of $\sim60^\circ$ to the plane of the sky. Indeed, the two clusters have slightly different redshifts, with A3667 being closer to the observer than A3651. The general properties of the two clusters are listed in Table~\ref{tab:clusters}. For consistency, we adopted the same values used in the X-ray analysis \citep{Dietl24}, which were taken from the Meta-Catalogue of X-ray Detected Clusters of Galaxies \citep[MCXC;][]{Piffaretti11}. However, we note that both clusters have updated redshifts from \citet{Lauer14}.

A3667 is one of the most X-ray luminous local galaxy clusters, and is well studied across the whole electromagnetic spectrum. It shows an elongated spatial distribution of galaxies and thermal gas in the south-east (SE) to north-west (NW) direction, which -- together with the observations of X-ray surface brightness jumps and substructures -- indicates ongoing merger activity along the SE-NW axis \citep[e.g.][]{Vichilin01,Owers09,Sarazin16,Omiya24}. In the radio band, A3667 hosts two bright radio relics, which indicates the presence of shock waves propagating along the same merger axis \citep{Rottgering97,Hindson14,deGasperin22}. A3651 lies at a projected distance of about 13 Mpc from A3667, in the NW direction. No detailed study of this cluster has been performed. 

\citet{Dietl24} note the presence of a foreground cluster located within the filament, named S0840 \citep[$z$=0.0148 and $R_{500}=15.6\arcmin$,][]{Piffaretti11}. This is possibly part of the Pavo-Indo supercluster, a massive aggregate of galaxy clusters at $z=0.015$. However, with a detailed analysis, the authors exclude the contribution of this system to the X-ray emission of the filament outside 4$\arcmin$ from the cluster's centre and excluded this region from their analysis.

\section{Methods}
\label{sec:methods}

ASKAP observed the A3667/3651 system for 10 hours in band 1 (800 - 1088 MHz) on $3$ August 2023, within Scheduling Block (SB) 51818. We downloaded the EMU total intensity image of this SB from the Commonwealth Scientific and Industrial Research Organization (CSIRO) ASKAP Science Data Archive (CASDA)\footnote{\url{https://data.csiro.au/domain/casdaObservation}}. The image was processed with the ASKAPsoft pipeline (version 1.11.1) and convolved to a resolution of $15\arcsec$ ($\sim17$ kpc at the system redshift). A zoom-in on the region around the A3667/3651 system is shown in Fig.~\ref{fig:EMU}. The root-mean-square noise of the image is 25 $\mu$Jy/beam. This is generally higher than the typical noise in  Stokes $Q$ and $U$, which is 17 $\mu$Jy/beam. Residual imaging artefacts are visible around a bright and extended radio galaxy (PKS 1954-55, also known as LEDA 63899), which is located in the vicinity of A3651 at $z$=0.058 \citep{Stein96}. This source is also bright in X-rays and, as pointed out by \citet{Dietl24}, might be the central galaxy of a galaxy group within the filament structure. Another extended radio galaxy with a remarkable X-shaped morphology is also visible in the field, to the north-east of A3667. This is LEDA 64440, at $z=0.061$ \citep{Koss22}. 

The Stokes $I$, $Q$, and $U$ frequency cubes were also downloaded from CASDA, together with the source catalogue produced from the Stokes $I$ cube by the Selavy software \citep{Whiting12}. The cubes were convolved to a common resolution ($20\arcsec$, which is $\sim22$ kpc at the system redshift). Q and U cubes were then corrected for ionospheric RM effects\footnote{\url{https://frion.readthedocs.io/en/latest/}}, and finally ingested by the POSSUM single-SB 1d-pipeline, which will be presented in a future paper (Van Eck et al., in prep.). This processing differs only slightly from that used for the full POSSUM survey \citep[as described by][]{Gaensler25}, mainly in that it uses data from a single SB rather than working with the tiles produced by mosaicked SBs. Here, we summarise the main steps of this single-SB pipeline:
\begin{itemize}
    \item We extracted $I$, $Q$, $U$ spectra from each source in the catalogue.
    \item We computed the foreground diffuse emission spectrum around each source, if present, and subtracted it from the source spectrum.
    \item We performed RM-synthesis on the spectra using RM-Tools\footnote{\url{https://github.com/CIRADA-Tools/RM-Tools}} \citep{Purcell20,VanEck26} to derive the Faraday dispersion function (FDF), or Faraday spectrum.
    \item We determined the peak of the Faraday spectrum with a parabolic fit to the FDF and derived polarisation parameters, such as RM, polarised intensity ($P$), polarisation fraction ($p=P/I$), and their uncertainties. The RM uncertainty was derived as the full width at half maximum (FWHM) of the FDF of each source, divided by two times the signal-to-noise ratio of the polarisation detection.
    \item We merged this information into the source catalogue.
\end{itemize}

The output catalogue follows the standard convention for RM catalogues introduced by \citet{VanEck23}, and polarisation spectra were also saved in a separate table.

\subsection{Quality RM selection}
\label{sec:qualitycuts}

   \begin{figure}
        \centering
        \includegraphics[width=0.9\linewidth]{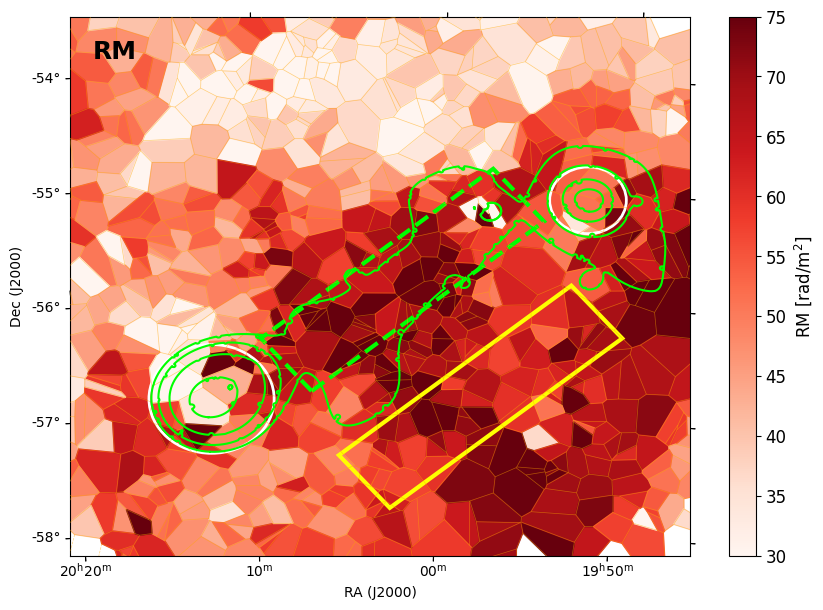}
        \includegraphics[width=0.9\linewidth]{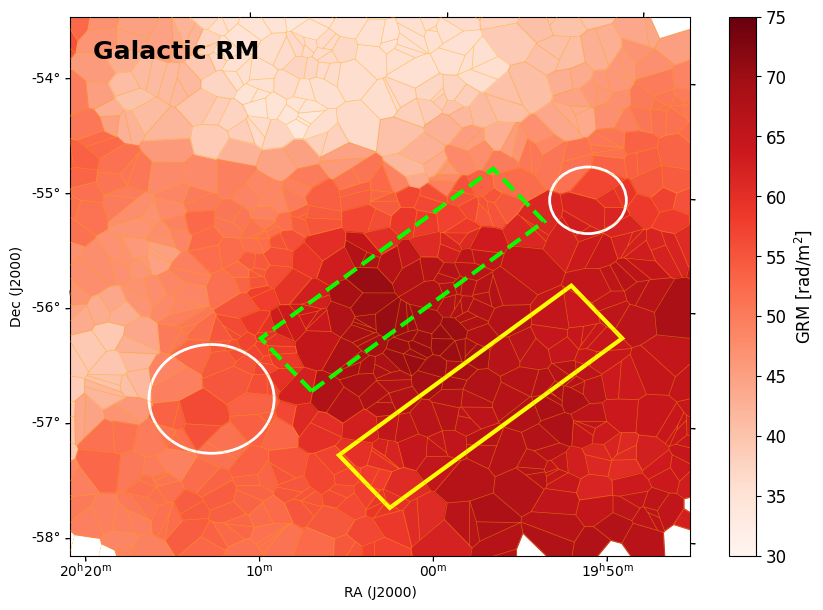}
        \includegraphics[width=0.9\linewidth]{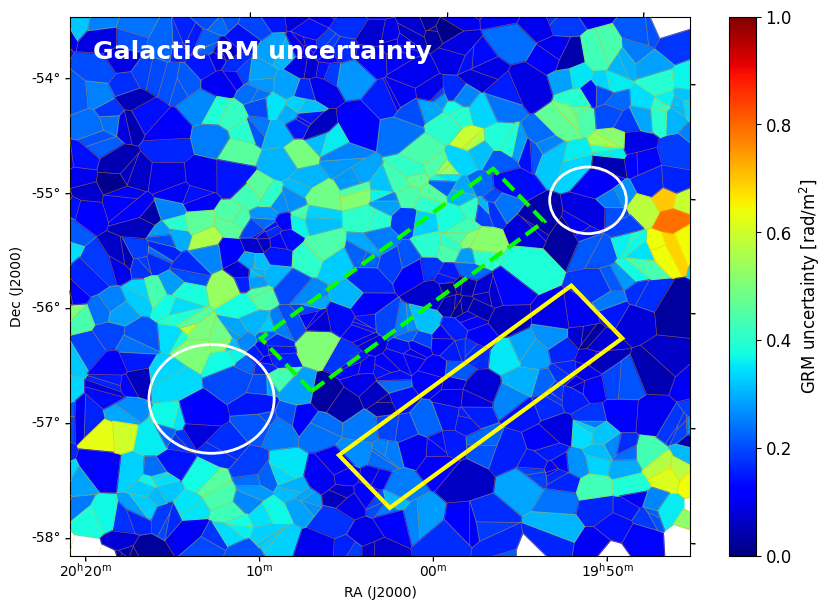}
        \caption{Voronoi visualisation of the polarised source catalogue. Top: Original RM catalogue. Centre: GRM computed with the annulus method. Bottom: GRM uncertainty at the position of each source (see Sect.~\ref{sec:GRM}). Each polygon represents a source in the catalogue. Its shape depends on the position of neighbouring sources, while its colour is described by the colour bar. The circles mark galaxy clusters as in Fig.~\ref{fig:EMU}, and X-ray contours from \citet{Dietl24} are also shown in the top panel. The dashed green rectangle marks the filament region, while the yellow rectangle marks the control region (defined in Sect.~\ref{sec:extgalRM}).}
        \label{fig:voronoi_RM-GRM}
    \end{figure}

The initial catalogue consists of 28989 sources, detected in total intensity. We performed several quality cuts in order to ensure the reliability of the polarised sources used for the analysis.
We thus excluded:

\begin{itemize}
    \item sources with signal-to-noise ratio in polarisation lower than 8. This is a standard threshold for Faraday rotation studies using RM synthesis. For POSSUM specifics (i.e. resolution in Faraday space of 58 rad m$^{-2}$ and maximum measurable RM of 8100 rad m$^{-2}$), this implies a detection with an equivalent Gaussian significance of 6.9$\sigma$ \citep{Hales12};
    \item sources with a low-quality total intensity spectral fit, identified by the pipeline parameter \texttt{IfitStat}>5;
    \item sources with fractional polarisation lower than $1\%$. This choice was motivated by the fact that, in this field, the average residual off-axis leakage from Stokes $I$ to $Q$ and $U$ is consistent with this value, as shown by the validation report\footnote{\url{https://ingest.pawsey.org.au/casda-support-files/validation/ASKAP/51818/AS203/POSSUM-validation-EMU_2006-55.SB51818.contcube/EMU_2006-55/index.html}};
    \item duplicate sources within a radius of $17\arcsec$, keeping only that with the highest signal-to-noise ratio relative to polarisation. The radius was chosen to be smaller than the image resolution ($20\arcsec$) to not exclude real double-lobed sources. Other polarised sources associated with artefacts around bright sources were manually discarded.
\end{itemize}

Following these quality cuts, we obtain a polarised source catalogue with 947 sources covering an area of about 30 deg$^2$. The catalogue is available at the CDS. This implies an RM density of $\sim32$ RM/deg$^2$, consistent with previous results using POSSUM pilot survey data \citep{Vanderwoude24}.

Polarised sources considered for this work are highlighted with grey circles in Fig.~\ref{fig:EMU}, where we also show the region of the filament, as detected by eROSITA, with the X-ray contour levels obtained by \citet{Dietl24}. We note that no polarised sources are located within the region contaminated by the X-ray emission of the foreground S0840 cluster. 

\subsection{Estimate of the Galactic RM}
\label{sec:GRM}

The RM values of the sources in the final catalogue were visualised using a Voronoi tessellation in the top panel of Fig.~\ref{fig:voronoi_RM-GRM}. The median RM across the entire SB is +53 rad m$^{-2}$, with a standard deviation of 17 rad m$^{-2}$. In the same region, the Galactic RM (GRM) estimated by \citet{Huts21} has a median value of +41 rad m$^{-2}$ and a standard deviation of 5 rad m$^{-2}$. The comparison indicates that the observed RM signal is largely dominated by the Galactic foreground contribution. However, our dense RM grid reveals coherent variations on angular scales below $1^\circ$, which were not captured by previous GRM estimates based on sparser RM catalogues.

Despite its relatively high Galactic latitude ($b=-33^\circ$), the Abell 3667/3651 system is observed along a line of sight towards the inner Galaxy, so its radiation still passes through a significant column of Galactic foreground material before reaching the observer. Moreover, it lies in a region characterised by enhanced Galactic diffuse polarised emission, as revealed by single-dish surveys \citep{Carretti13,Vidal15,West21}. Magnetic fields and thermal electrons in the intergalactic medium are expected to introduce significant Faraday rotation with fluctuations on large angular scales. Given that the X-ray-detected filament extends over more than $2^\circ$, it is crucial to assess the extent to which GRM may affect our measurements.

We estimated the GRM from the POSSUM RM using the annulus method introduced by \citet{Anderson24}. In this approach, the GRM at the position of each source is computed as the median RM of the nearest 40 sources located outside a given exclusion radius. The exclusion radius is introduced to avoid incorporating extragalactic spatial coherence in the RM distribution into the foreground GRM estimate. 

In our case, the field contains extragalactic RM structures spanning different angular scales, while the GRM also varies on multiple characteristic scales. We therefore adopted the following procedure:
\begin{itemize}
    \item For each source, we considered 40 exclusion radii equally spaced between 0.2$^\circ$ and 0.6$^\circ$. For each radius, all sources within the corresponding angular distance from the central source were excluded. This range corresponds to physical scales from $\sim0.8$ Mpc to $\sim2.4$ Mpc at the redshift of the system, comparable to $R_{500}$ of Abell 3651 and to the transverse extent of the filament, respectively.
    \item For each exclusion radius, we selected the 40 nearest sources outside the excluded region and computed their median RM value. Depending on the adopted exclusion radius, the farthest source used in the calculation lies between 0.5$^\circ$ and 1.3$^\circ$ from the central source (i.e. impact parameters of $\sim2$--$5$ Mpc).
    \item The GRM at the position of each source was then computed as the average of the 40 resulting median RM estimates.
    \item The uncertainty on the GRM was estimated as the standard deviation of these 40 values divided by $\sqrt{40}$.
\end{itemize}

The resulting GRM map and its uncertainty are shown in Fig.~\ref{fig:voronoi_RM-GRM}. The median value across the entire SB is +52 rad m$^{-2}$ with a standard deviation of 10 rad m$^{-2}$. This map can be regarded as a smoothed representation of the RM distribution, constructed under the assumption that the large-scale structures are dominated by Galactic foreground emission, without any a priori knowledge of their true origin. By adopting a range of exclusion radii, we assessed the sensitivity of the GRM reconstruction to the arbitrary choice of this parameter and quantify the uncertainty associated with it. The resulting uncertainties range from 0.09 to 0.95 rad m$^{-2}$ and increase in regions exhibiting strong spatial RM gradients, where the inferred GRM is more sensitive to the adopted smoothing scale.

Following this approach, we can thus estimate the residual RM (RRM),

\begin{equation}
    \rm RRM = RM - GRM~.
    \label{eq:RRM}
\end{equation}
Its uncertainty is the quadrature sum of the uncertainty on RM and on the GRM. 

\section{Results}
\label{sec:results}

Using the POSSUM polarised source catalogue, we created a dense RM grid that can be used to investigate the magnetic field in the A3667/3651 filament. We also derived an estimate of the Galactic contribution to the total RM at the position of each source. The large-scale variations visible in the RM map (Fig.~\ref{fig:voronoi_RM-GRM}) are also clearly present in the reconstructed GRM, which increases towards the south-western quadrant of the field and reaches maximum values of $\sim+70$ rad m$^{-2}$ in correspondence with the filament region. The spatial coincidence between the GRM enhancement and the filament, together with the fact that the characteristic angular scale of the GRM variations is comparable to the extent of the filament itself, motivates a careful consideration of additional Galactic tracers. In this section, we first compare the POSSUM RM grid with other Galactic tracers, derived with single-dish instruments (Sect.~\ref{sec:galtracers}). We then investigate the extragalactic RM contribution, focusing on the region of the filament (Sect.~\ref{sec:extgalRM}), and derive an estimate of its magnetic field (Sect.~\ref{sec:magestimate}).

\subsection{Comparison with Galactic tracers}
\label{sec:galtracers}

        \begin{figure*}
    \centering
    \stackinset{l}{23pt}{t}{7pt}{
        \includegraphics[width=0.14\textwidth]{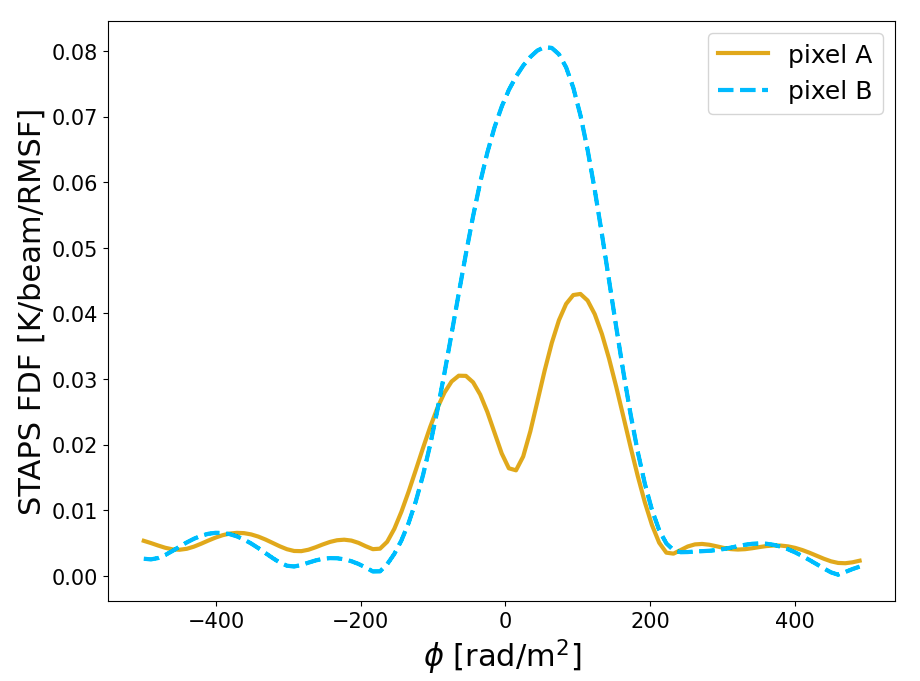}
    }{
        \includegraphics[width=0.48\textwidth]{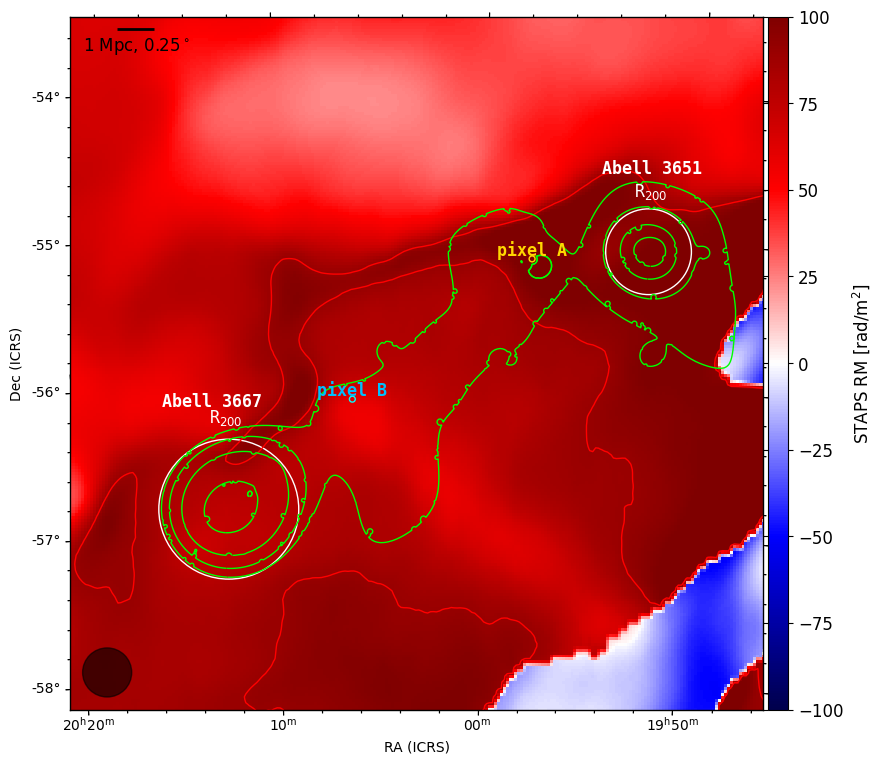}
    }
            \includegraphics[width=0.48\linewidth]{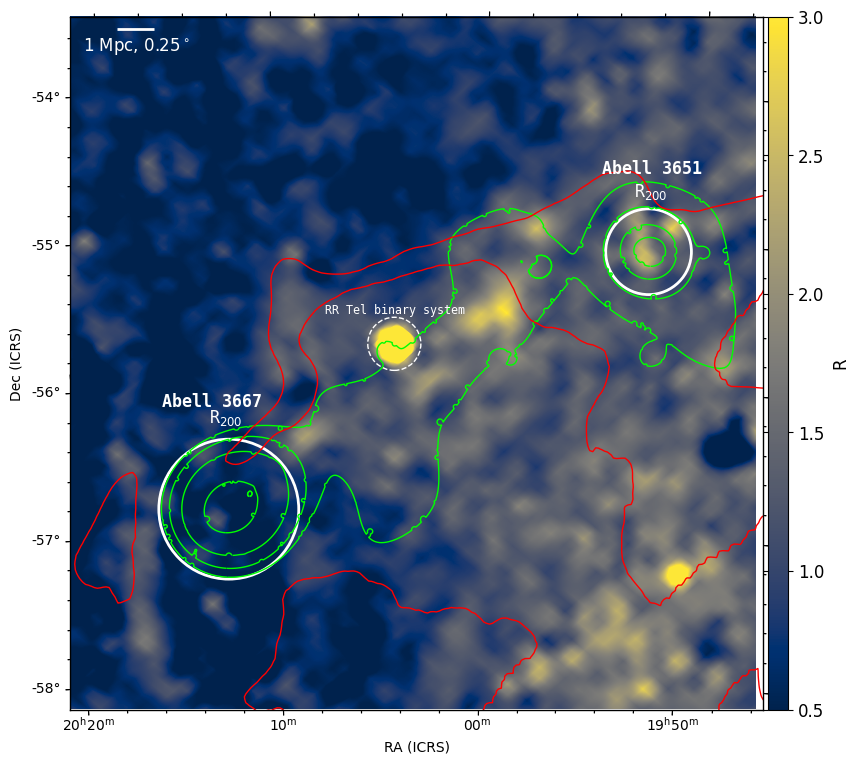}

    \caption{GRM tracers. Left panel: STAPS RM image \citep{Sun25}. A red contour shows the value of RM = +90 rad m$^{-2}$, highlighting the Galactic filament to the north of the extragalactic one. The regions and green contours are the same as in Fig.~\ref{fig:voronoi_RM-GRM}. Two reference pixels are marked, and their FDFs obtained by STAPS are shown in the inset panel. Right panel: H$\alpha$ image \citep{Finkbeiner03}. The red contour is identical to that in the left panel, while the cluster regions and green contours are the same as in Fig.~\ref{fig:voronoi_RM-GRM}. The RR Telescopii binary system is also marked with a dashed circle.}
    \label{fig:galtracers}
\end{figure*}

The Galactic Faraday rotation affecting the polarised radio emission produced within the Milky Way is also mapped by the Southern Twenty-centimetre All-sky Polarization Survey \citep[STAPS,][]{Sun25}, conducted with Murriyang, the Parkes 64-m telescope. The survey covers the southern sky, providing full-polarisation data in the frequency range 1.3–1.8 GHz, and delivers peak polarised intensity and Faraday depth maps derived through RM synthesis. All STAPS products are smoothed to a common angular resolution of $20\arcmin$. The STAPS RM in the filament region is shown in the left panel of Fig.~\ref{fig:galtracers}.

The STAPS map displays an RM distribution broadly consistent with that observed in the POSSUM RM (and GRM) maps. However, it also reveals structures with RM values exceeding $+90$ rad m$^{-2}$ and some regions with negative RMs, which are not visible in the POSSUM RM grid. In particular, a high-RM filament appears to connect the two Abell clusters. It is unresolved at the STAPS resolution, implying a width smaller than $20\arcmin$, and overlaps with the north-western part of the extragalactic filament, at the position of the PKS 1954-55 radio galaxy, while it extends beyond to the north-east. This filament is a segment of the northern boundary of a larger RM structure enclosing the area with RA from $\sim$20h30m to $\sim$19h40m and Dec from $\sim$-59$^\circ$59$\arcmin$ to $\sim$-54$^\circ$53$\arcmin$, as can be seen from the all-sky RM image from STAPS. Within this structure, another filament is visible starting from 20h08m -56$^\circ$00$\arcmin$ and running from the north-east to the south-west, which instead shows RM values of $\sim+65$ rad m$^{-2}$ in both STAPS and POSSUM.

Differences between the GRM measured by STAPS and that inferred from POSSUM are expected because the former probes only the Galactic Faraday rotation in front of the diffuse polarised Galactic emission, whereas the latter measures the total GRM integrated along the entire line of sight. Under the simplified assumption of uniform thermal electron density and magnetic-field strength and direction throughout the Galaxy (the Burn slab model; \citealt{Burn66}), the RM derived from STAPS is expected to be approximately half of that measured by POSSUM.

\begin{figure}
        \centering
        \includegraphics[width=0.9\linewidth]{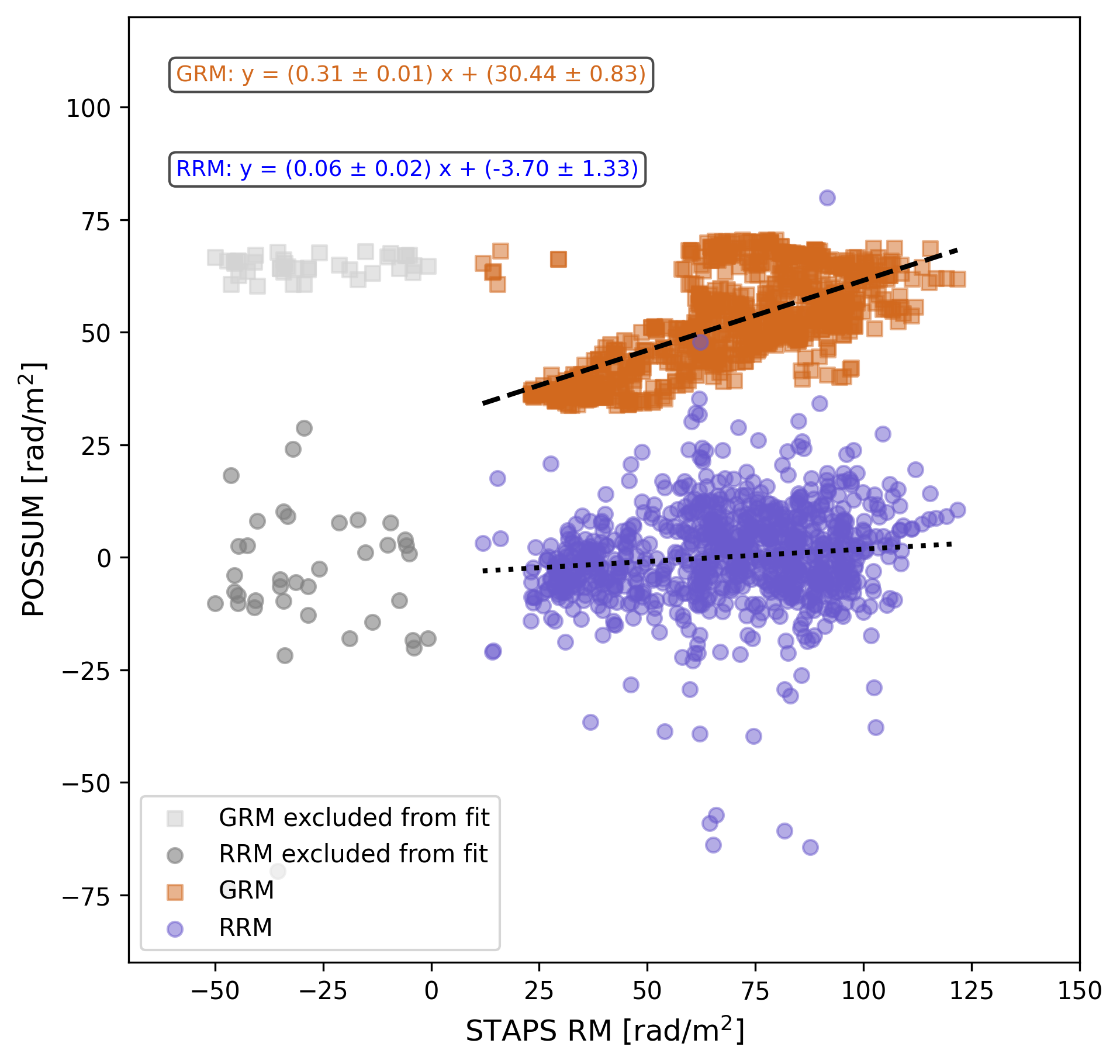}
        \caption{Scatter plot of the GRM and RRM computed from POSSUM vs the STAPS RM in the entire SB 51818. A linear fit between the two plotted quantities is also shown for both the GRM and RRM. Sources with negative STAPS RM values excluded from the fit are shown in grey.}
        \label{fig:STAPS-POSSUM}
    \end{figure}

We compare the RM measured by STAPS and the GRM estimated by POSSUM in Fig.~\ref{fig:STAPS-POSSUM}. POSSUM GRMs are systematically lower than the corresponding STAPS values. The result of a linear fit between the two quantities is also shown in the figure. For the fit, we excluded sources with negative STAPS values, as these clearly probe a different line of sight from that intersecting the extragalactic filament. The observed trend is inconsistent with the expectation from the simple Burn slab model. This discrepancy indicates that the assumptions underlying the Burn slab model are not valid in the present case and that STAPS RMs cannot be used as a reliable proxy for the total GRM in this region.

This interpretation is supported by the complex FDFs observed in the STAPS data. Two representative examples are shown in the inset of Fig.~\ref{fig:galtracers}, left plot. In the high-RM filament (pixel A), the RM map computed using the main peak of the spectrum traces a dominant positive component. However, an additional negative component is also present along the line of sight. The combined contribution of these components results in a reduced net RM, consistent with the lower values measured by POSSUM, and also results in a lower linear polarisation computed from the main peak, which is observed in STAPS along the filament. In this region, the two components are clearly separated in Faraday depth. In contrast, in other regions (e.g. pixel B, located in the perpendicular low-RM filament), the second component is near Faraday depth zero and only one peak is visible in the FDF after RM cleaning, showing a value closer to the POSSUM RMs. The negative component becomes instead dominant in the south-west. 

This comparison indicates that the Galactic foreground RM in the considered region is highly complex and likely dominated by local polarised structures, rather than by a smooth distribution of thermal electron density and magnetic field along the line of sight. Therefore, while our GRM estimate based on the annulus method is the best representation that we can have of the total Faraday effect of the Galaxy, we are still subject to the uncertainty on the scales of GRM structures that we would like to subtract.

Another useful comparison involves tracers of the Milky Way thermal gas. These include H$\alpha$ and H I line observations, which probe the warm ionised medium ($\sim8000-10000$ K) and the cold to warm neutral medium ($\sim100-5000$ K), respectively. We do not find a significant spatial correlation between the RM distribution and the H I data from \citet{HI4PIcoll16}. In contrast, a clear correspondence is observed between RM and H$\alpha$ emission, as shown in the right panel of Fig.~\ref{fig:galtracers}, where we present the H$\alpha$ map from \citet{Finkbeiner03}.
In particular, the region of enhanced RM observed by POSSUM in the south-western quadrant of the field overlaps with an area of increased H$\alpha$ emission and is also surrounded by the highest RM values detected in STAPS. In addition to the bright, compact H$\alpha$ source associated with the stellar system RR Telescopii, bright diffuse emission features are visible in the northern part of the extragalactic filament. This spatial coincidence suggests a possible Galactic contribution not only to the RM signal but also to the X-ray detection of the filament. This point is further discussed in Sect.~\ref{sec:discussX}.

We note that this system also lies in projection near the southern edge of the candidate supernova remnant (SNR) G354.0$-$33.5 \citep{Fesen21}. This is an extremely extended SNR, centred around 20h16m $-45^\circ50\arcmin$, with a diameter of $\sim11^\circ$ and exhibiting prominent H$\alpha$ emission extending towards the south. However, the H$\alpha$ emission shown in Fig.~\ref{fig:galtracers} does not appear to be clearly associated with this structure. \citet{Fesen21} also reported that ROSAT data do not show any obvious X-ray counterpart of this SNR.

\subsection{Estimate of the extragalactic contribution}
\label{sec:extgalRM}

Having assessed the complexity of the Galactic foreground, we conclude that the most appropriate approach to isolate the extragalactic contribution is to subtract the GRM derived with the annulus method from the POSSUM RM grid (Eq.~\ref{eq:RRM}), while accounting for the associated uncertainties in the GRM estimate, which stem from the uncertainty on the angular scale of Galactic structures. At the same time, it should be noted that this approach could alter the extragalactic signal. This possibility is discussed in Appendix~\ref{app:A}, where we tested the effect of our GRM extraction method on a mock RRM distribution.

\begin{figure}
        \centering
    \includegraphics[width=\linewidth]{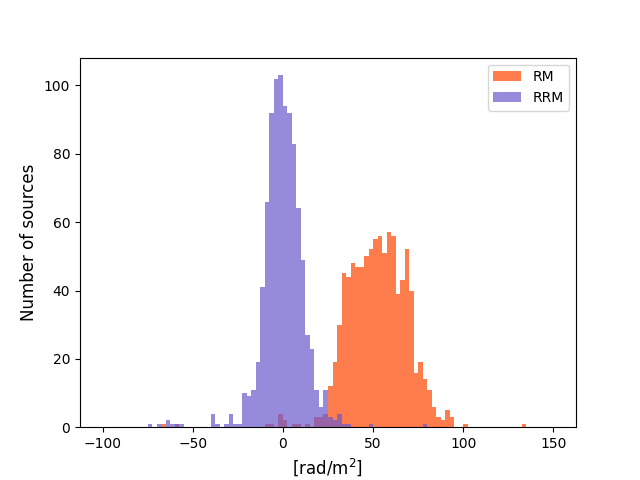}
        \caption{Histogram of the RM and RRM values in the entire SB 51818.}
        \label{fig:RMhisto}
    \end{figure}

The RRM represents the extragalactic contribution to the RM and is expected to be randomly distributed on large angular scales. Consistent with this expectation, the median RRM across the SB is zero, with a standard deviation of 13 rad m$^{-2}$. The RM and RRM distributions are compared in Fig.~\ref{fig:RMhisto}, showing that the annulus method effectively removes the average GRM contribution along each line of sight and reduces large-scale RM fluctuations. We further examine the relation between the RRM and the STAPS RM in Fig.~\ref{fig:STAPS-POSSUM}. The weakness of the correlation between the two quantities, indicated by the linear fit shown in the figure, further supports the conclusion that the RRM is largely independent of the Galactic foreground traced by STAPS.

   \begin{figure*}
        \centering
        \includegraphics[width=0.495\linewidth]{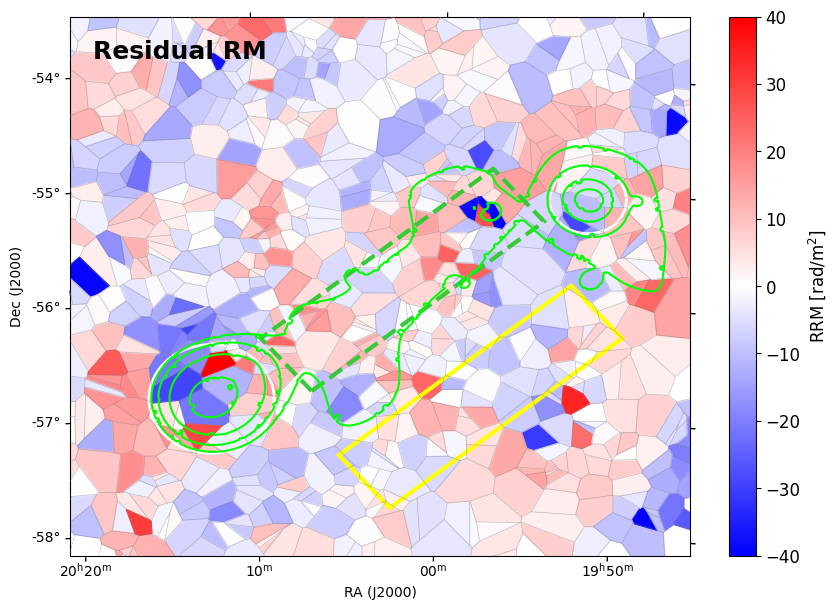}
        \includegraphics[width=0.495\linewidth]{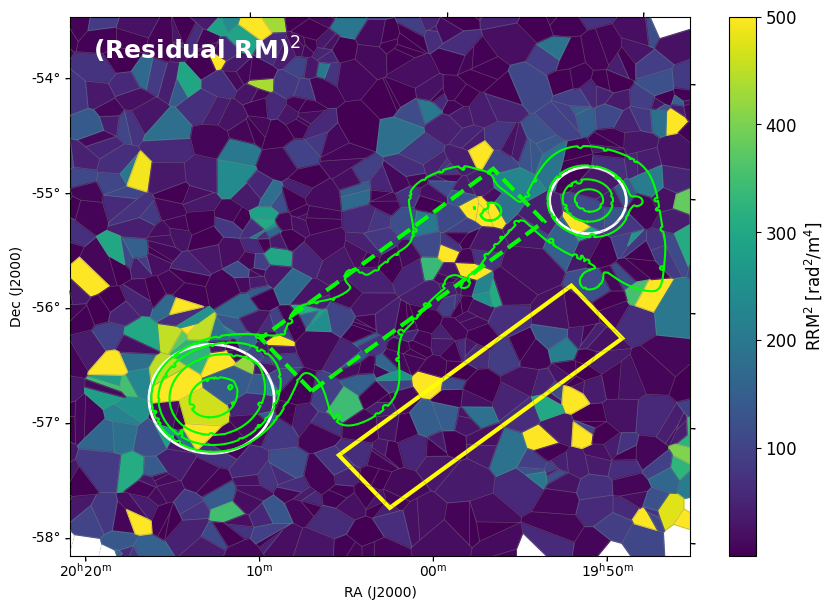}
        \caption{Voronoi visualisation of the polarised source catalogue in the region of the filament. Left: RRM computed with Eq.~\ref{eq:RRM}. Right: Its squared value. Each polygon represents a source in the catalogue. Its shape depends on the position of neighbouring sources, while its colour is described by the colour bar. The circles are the same as in Fig.~\ref{fig:EMU}, and X-ray contours from \citet{Dietl24} are also shown in the left panel. The dashed green rectangle marks the filament region, while the yellow rectangle marks the control region.}
        \label{fig:voronoi_RRM-RRM2}
    \end{figure*}

The Voronoi visualisations of the RRM map and its squared values are shown in Fig.~\ref{fig:voronoi_RRM-RRM2}. These maps reveal enhanced RRM values within the most massive cluster, A3667. We also observe elevated RRM values beyond the cluster $R_{200}$, in the direction of LEDA 64440. The RRM$^2$ map further suggests that the enhancement around A3667 is spatially offset, appearing displaced from the cluster centre. A similar behaviour has been reported in the case of the Fornax cluster \citep{Anderson21}.

Four sources with high RRM$^2$ values are also detected near the bright radio galaxy PKS 1954-55, which is part of the filament, suggesting the possible presence of a denser environment along the filament at this position, as also noticed by the X-ray analysis \citep{Dietl24}. We carefully inspected these high RRM$^2$ sources to verify that imaging artefacts do not cause them. One is indeed in the northern inner lobe of PKS 1954-55, which is known to be polarised and to show high negative RM values \citep{Morganti99}. Two are background objects, according to the photometric redshift derived from the Legacy Imaging Surveys Data Release 8 \citep{Duncan22}, and their RM could indeed be affected by the X-ray clump observed in the south-west of PKS 1954-55. The last one is a foreground edge-on spiral galaxy, ESO 185-G031 at $z=0.016$, found to have a large-scale galactic wind along the minor axis \citep{Fogarty12}, which could affect its polarisation properties, although this is not connected with the filament. We note that no RRM enhancement is seen around the foreground S0840 cluster, further confirming that its $R_{500}$ was probably overestimated, as previously claimed by \citet{Dietl24}.

Although an excess RRM is not obvious in the region of the filament from Fig.~\ref{fig:voronoi_RRM-RRM2}, it can be assessed through a statistical analysis. To derive statistical properties of the RRM in the filament, we defined two regions of interest, as shown in Figs.~\ref{fig:voronoi_RM-GRM} and ~\ref{fig:voronoi_RRM-RRM2}:

\begin{itemize}
    \item the filament region, defined based on the X-ray analysis by \citet{Dietl24}, in particular tracing the $2.3^\circ\times0.6^\circ$ ($\sim9.3\times2.4$ Mpc) rectangle used for their spectral analysis;
    \item a control region, of the same size as the filament region and displaced in the SW direction. This was chosen to lie in the region with a foreground Galactic contribution as similar as possible to that in front of the filament.
\end{itemize}

We have a total of 56 polarised sources in the filament region. We searched for optical counterparts of the sources seen through the filament in the Simbad archive\footnote{\url{https://simbad.cds.unistra.fr/simbad/}} and cross-matched their position with the photometric redshift catalogue of \citet{Duncan22}. Most of these sources reside behind the Abell 3667/3651 system and can be used to estimate its magnetic field, while we removed from the filament sample two low-redshift galaxies (2MASX J19582676-5535048 and ESO 185-G031 at $z=0.016$).

\begin{table}[]
    \centering
 \caption{RRM statistics in the considered regions.}
    \small
    \begin{tabular}{ccccc}
        Region & $N_{\rm sources}$ & $\sigma_\text{MAD,RRM}$ & $\widetilde{\text{RRM}}$ & $\widetilde{\text{p}}$ \\
               &             & rad/m$^{2}$ & rad/m$^{2}$ & $\%$  \\
               \hline
         Filament & 54 & 9.4$\pm$1.9 & 5.6$\pm$1.2 & 3.8$\pm$0.8 \\
         Control & 46 & 7.1$\pm$1.2 & -0.6$\pm$1.0 & 4.2$\pm$0.8 \\
         Abell 3667 & 20 & 13.5$\pm$4.5 & 6.5$\pm$3.5 & 4.6$\pm$1.1  \\
    \end{tabular}
    \tablefoot{Column 1: Region name. Column 2: Number of polarised sources in the region. Column 3: Robust estimate of the RRM standard deviation (1.48$\times$MAD). Column 4: Median RRM. Column 5: Median fractional polarisation. Uncertainties were estimated via bootstrapping.}
    \label{tab:RM}
\end{table}

   \begin{figure}
        \centering
    \includegraphics[width=\linewidth]{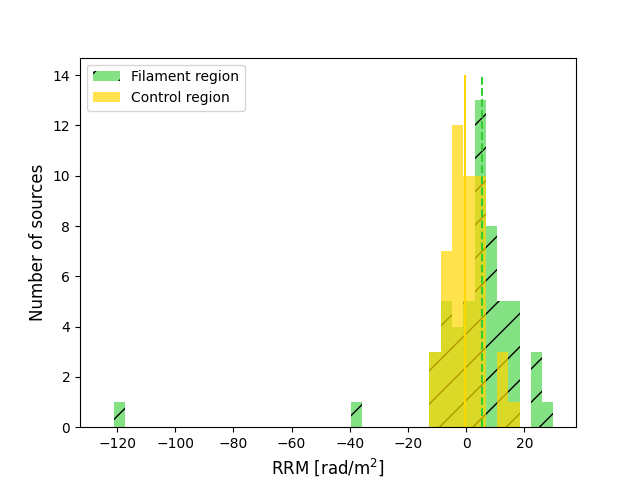}
        \caption{Histogram of the RRM values in the filament and the control region with median values marked with vertical lines.}
        \label{fig:RRMhisto}
    \end{figure}

The RRM distributions in the filament and control regions are shown in Fig.~\ref{fig:RRMhisto}. Since RRM assumes both positive and negative values and -- in the case of random magnetic field distributions -- are expected to average to zero, the RM dispersion, or standard deviation, is generally used to estimate magnetic field strength. However, RRM distributions displayed in Fig.~\ref{fig:RRMhisto} are not Gaussian, and outliers may strongly impact these estimates. We thus computed the robust estimate of the standard deviation of RM ($\sigma_\text{MAD,RM}$) as the median absolute deviation (MAD) multiplied by 1.48. This is generally used to have robust estimates of the standard deviation in the presence of outliers and with samples where a Gaussian distribution is poorly fitted. We then computed the uncertainty of $\sigma_\text{MAD,RRM}$ via bootstrapping, by taking the standard deviation of 10000 random resamples of the original distribution. Similarly, we computed the median RRM ($\widetilde{\text{RRM}}$) to estimate residual coherent RM structures, as well as the median polarisation fraction of sources ($\widetilde{\text{p}}$) to trace a possible depolarisation effect within the beam. Their uncertainties were estimated via bootstrapping. We note that the signed median is chosen as a representative value of the coherent RRM signal along the line of sight. The values are listed in Table~\ref{tab:RM}.

The same quantities were computed within $R_{200}$ of the Abell 3667 cluster. We note that only 20 background polarised sources are detected within this region, which, given the cluster’s angular extent, corresponds to an RM density of $\sim28$ deg$^{-2}$. This is slightly lower than the average density in the field ($\sim32$ deg$^{-2}$) but remains within the expected fluctuations in POSSUM fields \citep{Vanderwoude24}. Therefore, while depolarisation by the turbulent intra-cluster magnetic field may reduce the number of detectable polarised sources towards the cluster centre, the observed deficit is not statistically significant.

The RRM dispersion obtained in the control region is consistent with the value of $\sim 7$ rad m$^{-2}$ of intrinsic scatter of extragalactic sources observed at 1.4 GHz \citep{Oppermann15, Schnitzeler19}. The RRM dispersion is higher within the filament region, but still consistent within uncertainties, and increases further in the galaxy cluster region. A residual coherent RM signal is present both in the filament and in the cluster region, while it is absent in the control region, for which $\widetilde{\text{RRM}}$ is consistent with zero. The median fractional polarisation of sources is consistent among the three regions, indicating that there is not a major depolarisation effect on the scale of the beam (i.e. below 22 kpc).

\begin{figure}
        \centering
    \includegraphics[width=\linewidth]{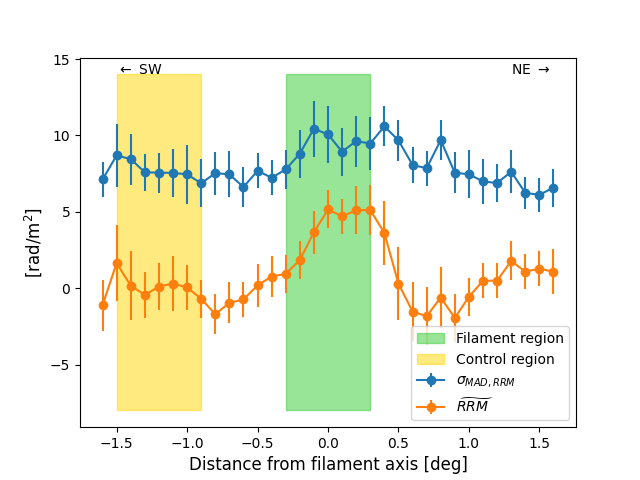}
        \caption{Median and $\sigma_{\rm MAD}$ of the RRM distribution in rectangular regions sliding from SW to the NE of the filament with a step of size $0.2^{\circ}$. The position of the filament and control regions selected for the analysis is also shown.}
        \label{fig:RRM_sliding}
    \end{figure}

To further investigate the RRM statistics across the field, we repeated the measurements of $\sigma_\text{MAD,RRM}$ and $\widetilde{\mathrm{RRM}}$ using sliding boxes with the same size as the filament and control regions, shifted by $0.2^{\circ}$ towards the south-west and north-east of the filament. The resulting statistics are shown in Fig.~\ref{fig:RRM_sliding}. The figure highlights a marginal enhancement in $\sigma_\text{MAD,RRM}$ and a clearer positive excess in $\widetilde{\mathrm{RRM}}$ in correspondence with the filament region. 

Compared to the filament region adopted in Table~\ref{tab:RM} (defined following the X-ray analysis), both enhancements appear slightly shifted towards the north-east. This could be due either to residual contamination from the GRM filament visible in the STAPS image (Fig.~\ref{fig:galtracers}) or to the presence of the PKS 1954-55 group. Using the STAPS data, we verified that residual RRM enhancements may indeed persist after the GRM subtraction in the region associated with the Galactic high-RM filament. However, these residuals are expected to be located predominantly to the north-east of the extragalactic filament, whereas Fig.~\ref{fig:RRM_sliding} shows an excess at the position of the filament itself.

\subsection{Magnetic field estimate in the filament}
\label{sec:magestimate}

The RRM scatter measured within the filament is marginally higher than that of the control region because they are both dominated by the intrinsic RM scatter of extragalactic sources. Assuming that both regions are affected by the same combination of intrinsic RM scatter from extragalactic sources and measurement uncertainties, we can isolate the filament contribution by subtracting in quadrature the RRM dispersion of the control region from that measured within the filament. Uncertainties are propagated using standard error propagation, and a $(1+z)^2$ correction factor is applied to both the value and its uncertainty to convert to the filament rest frame, adopting $z = 0.058$. We obtain

\begin{equation}
    \begin{split}
    \sigma_\text{RRM,excess} &= \sqrt{\sigma_\text{MAD,RRM,filament}^2-\sigma_\text{MAD,RRM,control}^2}(1+z)^2 \\
    &= 6.9\pm3.6 \ {\rm rad \ m^{-2}} .
    \end{split}
\end{equation}

The excess RRM dispersion reflects magnetic field and thermal electron density fluctuations on a broad range of scales, spanning between the minimum and maximum projected separations of the detected sources, i.e. between $\sim23\arcsec \simeq 26$ kpc and $\sim2.3^\circ \simeq 9.3$ Mpc, but also along the line of sight through the filament. In order to investigate the angular scales over which the RRM fluctuates, we computed the structure function (SF), which, following \citet{Haverkorn04}, is defined as

\begin{equation}
    {\rm SF}_{\mathrm{RRM}}(\delta\theta) =
    \left\langle
    \left[
    \mathrm{RRM}(\theta) -
    \mathrm{RRM}(\theta+\delta\theta)
    \right]^2
    \right\rangle .
\end{equation}

\begin{figure}
    \centering
    \includegraphics[width=0.85\linewidth]{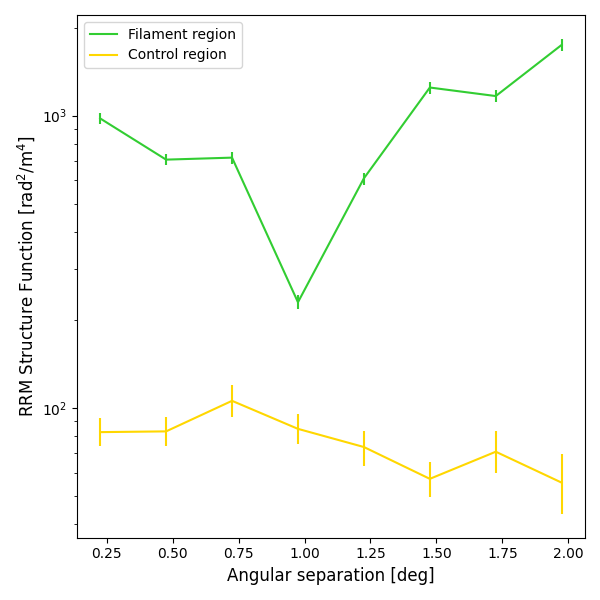}
    \caption{SF of the RRM distribution computed in the filament and control regions. The error bars indicate the 16$^{th}$ and 84$^{th}$ percentiles of the Monte Carlo realisations.}
    \label{fig:RRM_SF}
\end{figure}

We used the \texttt{StructureFunction} code\footnote{\url{Thomson, Alec (2024): StructureFunction. CSIRO. v1. Software. https://doi.org/10.25919/dkwn-mg50}}, which computes the SF and its uncertainties through Monte Carlo error propagation. The analysis was performed independently for the filament and control regions. Angular-separation bins were defined between $0.1^\circ$ and $2.1^\circ$, ensuring that each bin contained at least 50 RRM pairs. The resulting SFs are shown in Fig.~\ref{fig:RRM_SF}. The SF measured in the filament region is systematically higher than in the control region over the full range of sampled angular separations, consistent with enhanced RRM fluctuations in the filament on a physical scale between 400 kpc and 8.5 Mpc. This is consistent with that found by \citet{Gustafsson26} in the Abell 3391-3395 system.

The detected excess $\widetilde{\rm RRM}$ may trace a coherent Faraday rotation signal associated with the filament and not detected in the control region. This excess is significant with respect to the uncertainties associated with the GRM in the filament region (see Fig.~\ref{fig:voronoi_RM-GRM}), and may originate from coherent magnetic fields along the line of sight within the densest part of the filament. We note that $\widetilde{\rm RRM}$ is already computed after subtracting the median Galactic foreground contribution. Furthermore, the $\widetilde{\rm RRM}$ values measured outside the filament fluctuate around zero (Fig.~\ref{fig:RRM_sliding}), indicating the absence of a significant residual offset. We therefore interpret the measured $\widetilde{\rm RRM}$ as arising predominantly from the filament itself and use it as an estimate of the excess Faraday rotation due to the filament.

Correcting the filament $\widetilde{\rm RRM}$ and its uncertainty to the rest frame, we obtain

\begin{equation}
\widetilde{\text{RRM}}_\text{excess} =  6.3\pm1.3 \ {\rm rad \ m^{-2}} .
\end{equation}
The comparable values of $\widetilde{\rm RRM}_\text{excess}$ and $\sigma_\text{RRM,excess}$ suggest that the coherent and turbulent Faraday-rotating components within the filament contribute at similar levels to the observed RRM signal.

Magnetic fields in galaxy clusters are commonly studied through RM analyses and are known to be turbulent, with reversal scales ranging from a few kiloparsecs up to $\sim$1 Mpc \citep{Bonafede10,Stuardi21,Loi26}. These fluctuations are statistically described by a magnetic-field power spectrum, which in cosmological simulations typically peaks on scales of $\sim$100-200 kpc \citep[see e.g.][]{Vazza18,Dominguez19}. In filaments outside galaxy clusters, however, simulations predict that the peak of the magnetic-field power spectrum shifts towards larger scales, up to $\sim$1-3 Mpc, and that the RM signal may be dominated by the densest central region of the filament \citep{ChoRyu09,AkahoriRyu10}.

To derive magnetic field estimates, we used a first-order approximation, where we can describe the magneto-ionised medium with a single-scale model of a constant magnetic field with randomly aligned directions and uniform thermal electron density. In this case, the standard deviation of a distribution of N independent RM can be expressed as

\begin{equation}
\label{eq:sRM}
    \frac{\sigma_\text{RM}}{\rm rad\ m^{-2}} = \frac{812}{\sqrt{3}} \frac {n_e}{\rm cm^{-3}} \frac{B}{\mu G} \sqrt{\frac{L}{\rm kpc} \frac{\Lambda_c}{\rm kpc}},
\end{equation}
\noindent
where $\Lambda_c$ is the magnetic field reversal scale, $n_e$ is the constant thermal electron number density, $B$ is the constant modulus of the magnetic field ($B=\sqrt{B_x^2+B_y^2+B_z^2}=\sqrt3 B_\parallel$), and $L$ is the total depth of the medium \citep[see also][]{Gaensler01}. 

However, this formulation may overestimate the effective integration path, as it assumes a large number of field reversals along the line of sight. In reality, the RM signal may be dominated by the densest central region of the filament, with $L \sim \Lambda_c$ \citep{AkahoriRyu10}. In this limiting case, the coherent magnetic field can be estimated using the median RM value, and the previous expression reduces to

\begin{equation}
\label{eq:coherentRM}
    \frac{\text{RM}}{\rm rad\ m^{-2}} = \frac{812}{\sqrt{3}} \frac {n_e}{\rm cm^{-3}} \frac{B}{\mu G} \frac{\Lambda_c}{\rm kpc} .
\end{equation}

In the following, we discuss estimating the magnetic field strength using two limiting cases corresponding to (i) a random-walk regime, described by Eq.~\ref{eq:sRM} and (ii) a single-scale coherent slab, described by Eq.~\ref{eq:coherentRM}. These provide two estimates of the field strength, given the unknown magnetic coherence length in filaments. It is important to note that these are not independent estimates of different magnetic-field components (e.g. turbulent and ordered fields). Rather, they represent two alternative analytical descriptions used to infer the magnetic-field strength from the same RRM distribution under different assumptions about the characteristic coherence scale of the field. In reality, the magnetic field is expected to exhibit fluctuations over a broad range of spatial scales and is more appropriately described by a power spectrum than by a single coherence length. 

In both cases, complementary information on the thermal electron density distribution within the filament is required. Based on an X-ray spectral analysis and a surface brightness model using a truncated $\beta$-model, \citet{Dietl24} report a central baryon over-density of $\delta_0 = 215^{+86}_{-50}$, corresponding to a central thermal electron density of $5.5^{+2.2}_{-1.3} \times 10^{-5}$ cm$^{-3}$ at $z = 0.058$. In this model, the thermal electron density remains nearly constant at this value out to a radius of $\sim1$ Mpc, then decreases to $1 \times 10^{-5}$ cm$^{-3}$ at $\sim1.3$ Mpc. The authors caution that this estimate may be biased high due to the limited angular resolution of the data, which prevents the complete exclusion of all halo contributions. This concern is supported by a recent study that, through a more robust exclusion of point sources, measures a thermal electron density of $\sim10^{-5}$ cm$^{-3}$ in another inter-cluster filament \citep{Migkas25}. However, given that the filament between A3667 and A3651 is the first filament of such length to be detected in the X-rays, it is also plausible that its discovery could have been facilitated by it being intrinsically denser than the average population.

Following \citet{Pignataro25}, we adopted a Bayesian Markov chain Monte Carlo (MCMC) approach to explore the likelihood and reconstruct the posterior distribution of the two parameters, $B$ and $\Lambda_c$, in Eq.~\ref{eq:sRM} and Eq.~\ref{eq:coherentRM}. We included informed priors on the thermal electron density and on the total depth of the filament. For the density prior, we adopted a Gaussian distribution with mean $5\times10^{-5}$ cm$^{-3}$, corresponding to the average density within a radius of 1 Mpc in the model by \citet{Dietl24}, and standard deviation $2\times10^{-5}$ cm$^{-3}$. The total depth was fixed to $L = 4$ Mpc, assuming a cylindrical geometry for the filament with a radius of $\sim1$ Mpc and an inclination of approximately 60$^\circ$ along the line of sight.

In the random-walk regime, the likelihood function of detecting the observed $\sigma_{RRM,\rm excess}$ is

\begin{equation}
    \mathcal{L}(\sigma_{RRM,\rm excess} | \Lambda_c,B) \propto \exp \Bigl[ -\frac{1}{2} \Bigl( \frac{\sigma_{RRM,\rm excess} - \sigma_\text{RM}(\Lambda_c, B)}{\rm err_{\sigma_{RRM,\rm excess}}} \Bigr)^2 \Bigr]
    \label{eq:likelihood_sRM}.
\end{equation}
 
For the free parameters, we assumed flat priors in the ranges $0 < B < 5$ $\mu$G and $25 < \Lambda_c < 400$ kpc. The upper edge of the range of magnetic field strengths was chosen to be comparable to the value measured at the centre of galaxy clusters \citep[e.g.][]{Bonafede10}, while the coherence-scale range was selected between the minimum scale sampled by the POSSUM sources and a maximum value of 400 kpc, ensuring at least $\sim10$ magnetic-field reversals along the line of sight. The distribution is then sampled with MCMC using \texttt{emcee}\footnote{\url{https://emcee.readthedocs.io/en/stable/}} \citep{Foreman13}.

\begin{figure}
        \centering
        \hspace{-25pt}\includegraphics[width=0.91\linewidth]{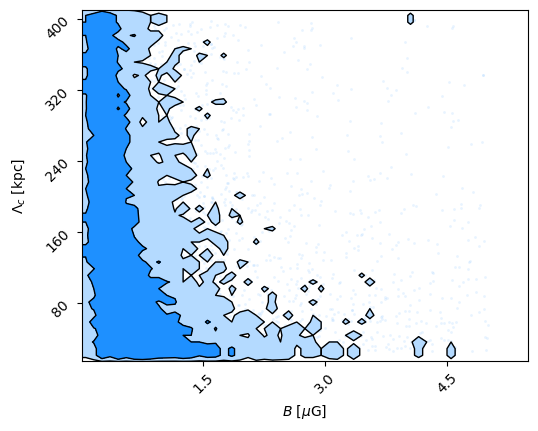}
       \includegraphics[width=0.83\linewidth]{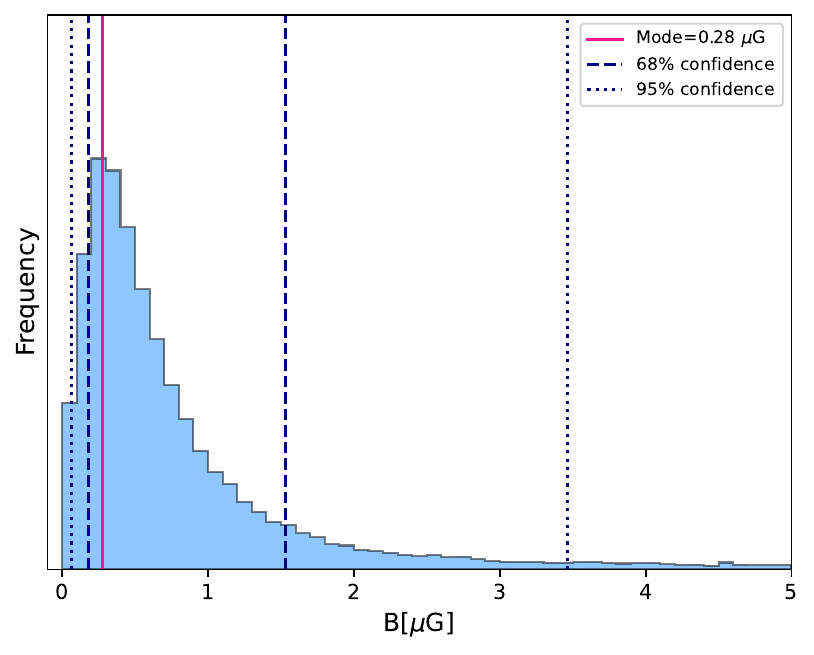}
        \caption{Posterior distribution from the MCMC sampling of $\sigma_{RRM,\rm excess}$. The model assumes a turbulent magnetised medium characterised by a single coherence scale, $\Lambda_c$ (Eq.~\ref{eq:sRM} and \ref{eq:likelihood_sRM}), and adopts informed priors on the total depth of the medium and thermal electron density. Top: Covariance of the posterior probability distribution for the coherence scale and magnetic field, with the $95\%$ confidence region filled in blue and the $68\%$ confidence region filled in light blue. Bottom: Marginalised magnetic field posterior probability distribution.}
        \label{fig:MCMC_sigmaRM}
    \end{figure}

The covariance of the posterior probability distribution of the two free parameters is shown in the top panel of Fig.~\ref{fig:MCMC_sigmaRM}. The shape of the covariance indicates the correlation between the two, which is clear from Eq.~\ref{eq:sRM}. The reversal scale $\Lambda_c$ is essentially unconstrained within the prior range. However, large values of magnetic field are excluded for all $\Lambda_c$. The magnetic field posterior distribution marginalised over $\Lambda_c$ is shown in the bottom panel of Fig.~\ref{fig:MCMC_sigmaRM} with its mode and $68\%$ and $95\%$ confidence levels. The most probable magnetic field computed from the RRM dispersion on small scales is thus: $B = 0.3^{+3.2}_{-0.2} ~ \mu$G within 95$\%$ confidence.

In the single-scale coherent slab assumption, the likelihood function of detecting the observed $\widetilde{RRM}_{\rm excess}$ is

\begin{equation}
    \mathcal{L}(\widetilde{RRM}_{\rm excess} | \Lambda_c,B) \propto \exp \Bigl[ -\frac{1}{2} \Bigl( \frac{\widetilde{RRM}_{\rm excess} - \text{RM}(\Lambda_c, B)}{\rm err_{\widetilde{RRM}_{\rm excess}}} \Bigr)^2 \Bigr]
    \label{eq:likelihood_coherentRM}.
\end{equation}
 
In this case, for the free parameters, we kept the same flat prior for the magnetic field, $0 < B  < 5$ $\mu$G, but we increased the values for the coherence length to $500 < \Lambda_c < 1500$ kpc. With this, we assumed that the RRM is dominated by the central densest region of the filament.

   \begin{figure}
        \centering
        \hspace{-25pt}\includegraphics[width=0.91\linewidth]{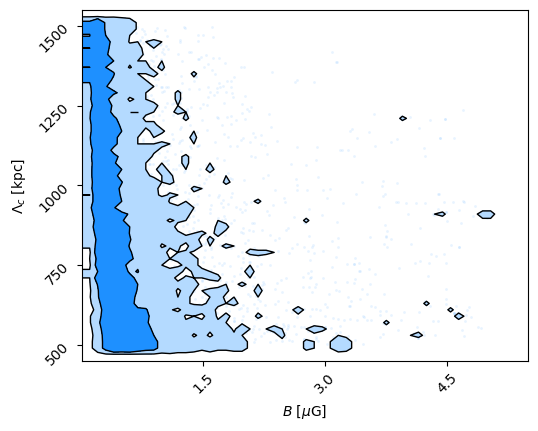}
        \includegraphics[width=0.83\linewidth]{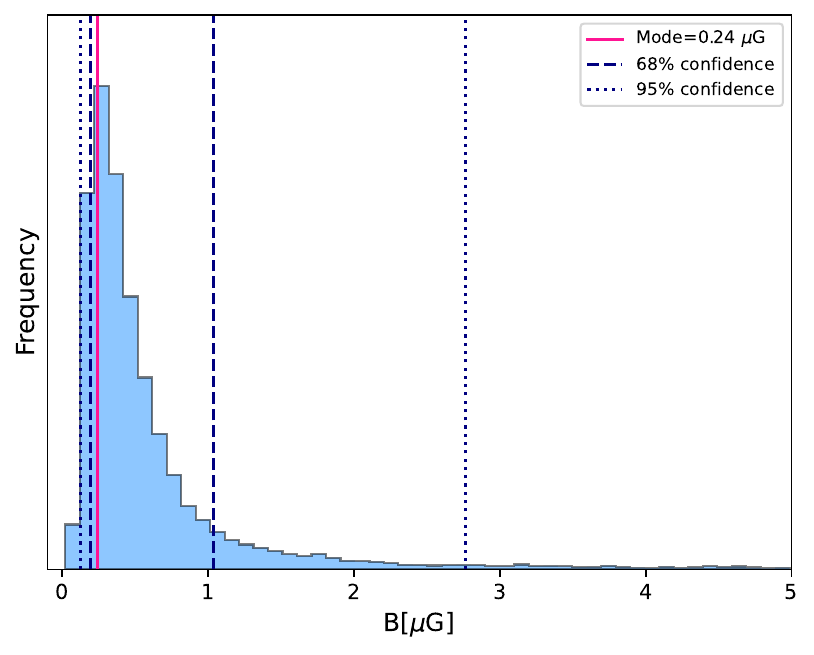}
        \caption{Posterior distributions from the MCMC sampling of $\widetilde{RRM}_{\rm excess}$. The model assumes a single-scale coherent slab (Eqs.~\ref{eq:coherentRM} and \ref{eq:likelihood_coherentRM}) and adopts an informed prior on the thermal electron density. Top: Covariance of the posterior probability distribution for the coherence scale and magnetic field, with the $95\%$ confidence region filled in blue and the $68\%$ confidence region filled in light blue. Bottom: Marginalised magnetic field posterior probability distribution.}
        \label{fig:MCMC_RM}
    \end{figure}

The covariance of the posterior probability distribution of the two free parameters is shown in the top panel of Fig.~\ref{fig:MCMC_RM} while the magnetic field posterior distribution marginalised over $\Lambda_c$ is shown in the bottom panel. The most probable magnetic field computed under the single-scale coherent slab assumption is thus: $B = 0.2^{+2.5}_{-0.1} ~ \mu$G within 95$\%$ confidence.

The two estimates are thus consistent, providing a $95\%$ confidence on the presence of a magnetic field with a strength larger than 0.1~$\mu$G in the filament. This result is discussed in the next section.

\section{Discussion}
\label{sec:discussion}   

\subsection{The importance of Galactic RM foreground for extragalactic magnetism science}

Our analysis highlights that Galactic Faraday rotation is likely to represent one of the main limiting factors for RM-grid studies of individual cosmic-web filaments, even at relatively high Galactic latitudes. In the A3667/3651 system, the angular scales of the GRM fluctuations are comparable to the projected extent of the filament itself, making the separation between Galactic and extragalactic contributions intrinsically difficult.

The comparison between the POSSUM RM grid and the STAPS diffuse-polarisation data demonstrates that the Galactic foreground in this field cannot be described as a smooth RM screen. Instead, the foreground is characterised by localised structures and by multiple Faraday components along the line of sight, producing strong spatial variations in RM on sub-degree scales. These structures are not fully captured by existing GRM reconstructions based on sparse RM catalogues \citep[e.g.][]{Huts21}, and can significantly affect the interpretation of weak extragalactic RM signals.

In this context, the annulus method used in this work provides a practical way to reconstruct the GRM directly from the local RM distribution measured by POSSUM. However, this approach also acts as a spatial high-pass filter and may partially suppress genuine extragalactic RM fluctuations on angular scales comparable to or larger than the exclusion radius adopted for the reconstruction. Our test, presented in Appendix~\ref{app:A}, suggests that the method is nevertheless able to recover excess RM fluctuations on the scale of the filament without introducing major biases.

These results suggest that future RM-grid studies with POSSUM and the Square Kilometre Array  \citep[SKA;][]{Dewdney09} will strongly benefit from combining compact-source RM catalogues with diffuse Galactic polarisation surveys, such as STAPS. Diffuse polarised emission contains crucial information on the Faraday complexity of the Galactic foreground and can help identify localised structures that may contaminate extragalactic RM measurements. Additional progress will also require denser RM grids, improved modelling of the Galactic magnetised interstellar medium, and statistical techniques capable of jointly reconstructing Galactic and extragalactic Faraday components across multiple angular scales \citep{Vacca26}.

Finally, our results indicate that detecting magnetic fields in individual filaments through RM-grid analyses remains extremely challenging with current gigahertz-frequency surveys. Even in the presence of a relatively nearby and X-ray-bright filament, the expected RM excess is comparable to both the intrinsic scatter of background-source RMs and the residual uncertainties associated with Galactic foreground subtraction. This suggests that robust detections of magnetic fields in individual filaments may require the combination of ultra-dense RM grids and lower-frequency observations

\subsection{Possible Galactic contribution to the X-ray detection of the filament}
\label{sec:discussX}

The complex Galactic foreground in the A3667/3651 system raises the question of whether the reported X-ray detection may be affected by Galactic contamination. In particular, the resemblance between the H$\alpha$ emission morphology and the northern edge of the filament X-ray contours, as well as the apparent spatial coincidence with the binary system RR Telescopii (see Fig.~\ref{fig:galtracers} right), warrants further consideration.

Regarding RR Telescopii, the alignment between the source position in the H$\alpha$ map and a local enhancement in the X-ray contours initially appears suggestive. However, inspection of the X-ray imaging confirms that RR Telescopii was fully excised from the eROSITA data processing in \citet{Dietl24}. The apparent enhancement in the X-ray contours corresponds to a distinct diffuse emission feature and is not associated with residual emission from the binary system.

A more challenging case is the possibility of diffuse Galactic foreground contamination on larger angular scales. The eROSITA maps show an increase in soft X-ray emission towards the west of A3651, i.e. in the direction of the Galactic plane, closely resembling the large-scale gradient observed in the H$\alpha$ emission. Since the thermal spectrum expected from the filament in the soft X-ray band overlaps significantly with the spectral characteristics of the Galactic foreground emission, a rigorous separation of the two components is challenging. To account for this uncertainty, \citet{Dietl24} modelled the Galactic foreground using two independent control regions located south-west and north-east of the filament, respectively. The south-western control region samples an area of enhanced Galactic soft X-ray and H$\alpha$ emission, while the north-eastern region probes comparatively fainter foreground emission. As expected, the inferred foreground normalisation differs between the two models. Nevertheless, the derived filament properties remain consistent within overlapping 1$\,\sigma$ uncertainties (Appendix D of \citealt{Dietl24}). This suggests that, while the precise surface brightness normalisation depends on the adopted Galactic foreground model, the presence of an extended filament excess itself is not driven by a particular foreground choice. However, the filament spectrum is extracted over a very large sky region ($2.3^\circ\times0.6^\circ$), across which spatial variations in the Galactic foreground are expected but not explicitly modelled. Consequently, a residual Galactic contribution, particularly in regions closer to the Galactic plane, cannot be ruled out. Such contamination would artificially enhance the inferred X-ray surface brightness and electron density of the filament. In that case, the inferred magnetic field strengths would be systematically underestimated.

\subsection{A first estimate of the magnetic field in an individual inter-cluster filament}

The results of Sect.~\ref{sec:magestimate} can be compared with previous estimates of magnetic fields in low-density cosmic environments. In the outskirts of galaxy clusters such as the Coma cluster, magnetic field strengths of $\sim$0.5~$\mu$G have been estimated at distances of $\sim$2 Mpc from the centre, where the density is $~\sim4\times10^{-5}$ cm$^{-3}$ \citep{Bonafede10}. Similarly, polarisation properties of radio relics - diffuse synchrotron sources observed in the outskirts of some merging galaxy clusters - indicate magnetic field strengths of 0.1-1 $\mu$G \citep{Stuardi22}. Observations of the bridge between Abell 399 and Abell 401 indicate values ranging from $\sim$0.1$\mu$G \citep{Govoni19} to $\sim$0.5–0.6~$\mu$G \citep{Brunetti20}, depending on the particle acceleration mechanism yielding the synchrotron emission, with some lower limits at 0.46~nG \citep{Balboni23}. The thermal electron density in this bridge is $3.4\times10^{-4}$ cm$^{-3}$ \citep{Fujita08} \citep[although a lower value of $9\times10^{-5}$ cm$^{-3}$ is reported by][considering projection effects]{Hincks22}. 

In our modelling, we used a thermal electron density of $\sim5\times10^{-5}$ cm$^{-3}$ \citep{Dietl24}. Our result is thus consistent with these estimates, allowing for a magnetic field strength that is typical of bridges and cluster outskirts and suggesting that, as well as in these environments, the magnetic field in the Abell 3667/3651 filament is already amplified by small-scale turbulence originating from structure formation processes. 

For energy considerations, we can compute the plasma beta, \begin{equation} \beta = \frac{8\pi n_e k_B T}{B^2}, \end{equation} where $k_B$ is the Boltzmann constant and $T$ the plasma temperature, in the central region of the filament using the thermal electron density and temperature derived from the X-ray analysis together with the magnetic field estimated in this work. Considering the uncertainties on the magnetic-field coherence scale, as well as a possible residual contribution from the complex Galactic foreground, we adopted a conservative value of $B=0.1\,\mu\mathrm{G}$, which is compatible with our estimates at the $95\%$ confidence level and favours larger magnetic-field coherence scales. For $n_e = 5.5\times10^{-5}\,\mathrm{cm^{-3}}$ and $T = 0.91\,\mathrm{keV}$, we obtain $\beta \sim 200$. This value is consistent with expectations for a high-$\beta$ plasma that has already been significantly processed by structure-formation activity and is less compatible with pristine cosmic-web filament environments, where substantially larger values ($\beta \gg 100$) are generally expected.

In cosmic web filaments, cosmological simulations predict that the dynamo process is still largely inefficient and that volume-averaged magnetic fields have strengths of the order of a few tens of nanogauss \citep{Vazza14a}. This prediction is consistent with recent RM stacking experiments, which inferred average magnetic-field strengths of 10-60 nG in local filaments \citep{Carretti22,Carretti23,Carretti25}. However, the density-weighted magnetic field relevant for Faraday-rotation studies may reach several tens to a few hundred nanogauss according to \citet{Ryu08}. Furthermore, the latter simulations predict a large scatter in magnetic-field strengths within filaments, suggesting that the small-scale dynamo is more efficient in the densest regions, although likely still operating in the linear growth regime. Since the RM stacking experiments were performed at megahertz frequencies, they are intrinsically less sensitive to sight lines exhibiting large Faraday-depth fluctuations, as these are more strongly affected by depolarisation. As a result, such measurements may preferentially probe the more diffuse regions of filaments. In contrast, our RM measurement is not subject to this selection effect and may therefore be dominated by the contribution from the denser central regions of the filament, where stronger magnetic fields are expected.

Furthermore, \citet{Dietl24} also pointed out that contamination from unsubtracted collapsed halos, small groups, and gas clumps could explain the discrepancy between their observations and predictions from cosmological simulations. This interpretation was recently supported by \citet{Migkas25}, who measured a thermal electron density of $\sim10^{-5}$ cm$^{-3}$ in the filament connecting the cluster pairs A3530/32 and A3528-N/S. If a similar thermal electron density characterises our filament, higher magnetic field strengths or longer correlation lengths would be required to account for the observed RM scatter. However, our RM measurements could likewise be influenced by the presence of gas clumps and halos, which make it difficult to use a homogeneous model for the thermal electron density. Notably, the enhancement in RM is most pronounced near the sub-group associated with PKS 1945-55 and in the vicinity of the more massive A3667 cluster. A denser RM grid would enable a more detailed investigation of the spatial variations in the magnetic field across the filament.

Another important aspect to consider is the broad range of magnetic field reversal scales explored in our modelling. While real magnetic fields are better described by a power spectrum, the peak of this spectrum corresponds to the scale where most of the magnetic energy is concentrated. In galaxy clusters, this peak is typically expected at scales of a few hundred kiloparsecs \citep[e.g.][]{Dominguez19}. However, in filaments, larger scales may dominate since turbulence is not fully developed as at the centre of galaxy clusters \citep{ChoRyu09,AkahoriRyu10,Vazza14a}. If this is the case, more stringent upper limits on the magnetic field strength apply, since for $\Lambda_c > 200$ kpc, magnetic fields stronger than 0.5~$\mu$G are strongly disfavoured (see upper panel of Fig.~\ref{fig:MCMC_sigmaRM}).

Finally, we note that with an RM grid consisting of approximately 50 sources, the uncertainties on $\sigma_{\mathrm{MAD,RM}}$ computed in both the filament and background regions are comparable to the average uncertainties on individual RM measurements in the POSSUM survey, i.e. 1–2 rad m$^{-2}$. Given that the intrinsic RM dispersion of sources at 1.4 GHz is typically around $\sim7$ rad m$^{-2}$ \citep{Oppermann15,Schnitzeler19}, increasing the number of sources by at least a factor of 15 would be required to reduce the statistical uncertainties and enable a reliable separation of the RM dispersion induced by the filament. Furthermore, the uncertainties related to the Galactic foreground contribution make the detection in this filament particularly challenging because the median RRM could be affected by residual Galactic signal. With the current observational capabilities, magnetic fields of the order of tens of nanogauss - as inferred from low-frequency RM studies in stacked superclusters and filaments \citep{Carretti25,Pignataro25} - remain elusive at gigahertz frequencies, as they would produce RM dispersion differences smaller than the present uncertainties. 

\section{Conclusion}
\label{sec:conclusions}

In this work, we performed the first RM-grid analysis of an individual X-ray-detected inter-cluster filament, using POSSUM observations of the system connecting the nearby galaxy clusters Abell 3667 and Abell 3651. The filament extends over a projected scale of $\sim 13$ Mpc and was recently detected in soft X-rays by eROSITA \citep{Dietl24}.

Using POSSUM, we constructed a dense RM grid with $\sim 32$ sources deg$^{-2}$ and estimated the GRM contribution through the annulus method with varying exclusion radii. We find that the Galactic foreground in this region is highly structured and characterised by significant RM fluctuations on angular scales comparable to the extent of the filament itself. Comparison with STAPS diffuse-polarisation data reveals that the foreground contains localised Faraday structures and multiple components along the line of sight, highlighting the complexity of GRM subtraction in cosmic-web studies.

After subtracting the reconstructed Galactic contribution, we investigated the residual extragalactic RM signal in the filament region. We measure a marginal excess RRM dispersion of
\[
\sigma_{\rm RRM,excess} = 6.9 \pm 3.6~{\rm rad~m^{-2}},
\]
together with a coherent residual signal with median
\[
\widetilde{\rm RRM}_{\rm excess} = 6.3 \pm 1.3~{\rm rad~m^{-2}}.
\]
The enhancement in RRM statistics is spatially associated with the filament region, although residual contamination from Galactic structures cannot be fully excluded.

Assuming a simplified single-scale magnetic-field model and adopting informed priors on the thermal electron density derived from the X-ray analysis, we constrained the magnetic field strength in the filament to the range $0.1$-$3.5~\mu$G within the 95\% confidence level. The most probable magnetic field value is $\sim 0.2$--$0.3~\mu$G, depending on the assumed coherence scale. Larger magnetic-field reversal scales favour lower magnetic-field strengths.

Our results provide the first direct constraints based on Faraday rotation measurements in an individual inter-cluster filament. The inferred magnetic-field strengths are consistent with expectations for magnetised gas in bridges and cluster outskirts and in the very central region of cosmic web filaments, where structure formation processes are already active.

This work also demonstrates that Galactic foreground subtraction is likely to become one of the dominant limitations for future RM-grid studies of individual targets. Combining dense RM catalogues with diffuse Galactic polarisation surveys and improved foreground modelling will therefore be essential for fully exploiting the capabilities of POSSUM and the SKA in studies of extragalactic magnetism.


\section*{Data availability}

The polarised source catalogue used in this work is publicly available at the CDS via anonymous ftp to \url{cdsarc.u-strasbg.fr} (130.79.128.5) or via \url{http://cdsweb.u-strasbg.fr/cgi-bin/qcat?J/A+A/}.


\begin{acknowledgements}
This scientific work uses data obtained from Inyarrimanha Ilgari Bundara / the Murchison Radio-astronomy Observatory. We acknowledge the Wajarri Yamaji People as the Traditional Owners and native title holders of the Observatory site. The Australian SKA Pathfinder is part of the Australia Telescope National Facility (https://ror.org/05qajvd42) which is managed by CSIRO. Operation of ASKAP is funded by the Australian Government with support from the National Collaborative Research Infrastructure Strategy. ASKAP uses the resources of the Pawsey Supercomputing Centre. Establishment of ASKAP, the Murchison Radio-astronomy Observatory and the Pawsey Supercomputing Centre are initiatives of the Australian Government, with support from the Government of Western Australia and the Science and Industry Endowment Fund. The POSSUM project (https://possum-survey.org) has been made possible through funding from the Australian Research Council, the Natural Sciences and Engineering Research Council of Canada, the Canada Research Chairs Program, and the Canada Foundation for Innovation. \\
SPO acknowledges support from the Comunidad de Madrid Atracción de Talento program via grant 2022-T1/TIC-23797, and grant PID2023-146372OB-I00 funded by MICIU/AEI/10.13039/501100011033 and by ERDF, EU. \\
DAL acknowledges support from the Universidad Complutense de Madrid and Banco Santander through the predoctoral grant CT25/24. \\
AB  acknowledges support from the ERC CoG $\vec{B}$ELOVED, GA N.101169773. \\
XS is supported by the National SKA Program of China (Grant No. 2022SKA0120101) and the National Natural Science Foundation of China (No. 12433006).
     
\end{acknowledgements}

\bibliographystyle{aa}
\bibliography{my_bib}

\clearpage
\onecolumn

\begin{appendix}

\FloatBarrier

\section{Test of Galactic RM estimate}
\label{app:A}

\begin{figure}[H]
        \centering
        \includegraphics[width=0.33\linewidth]{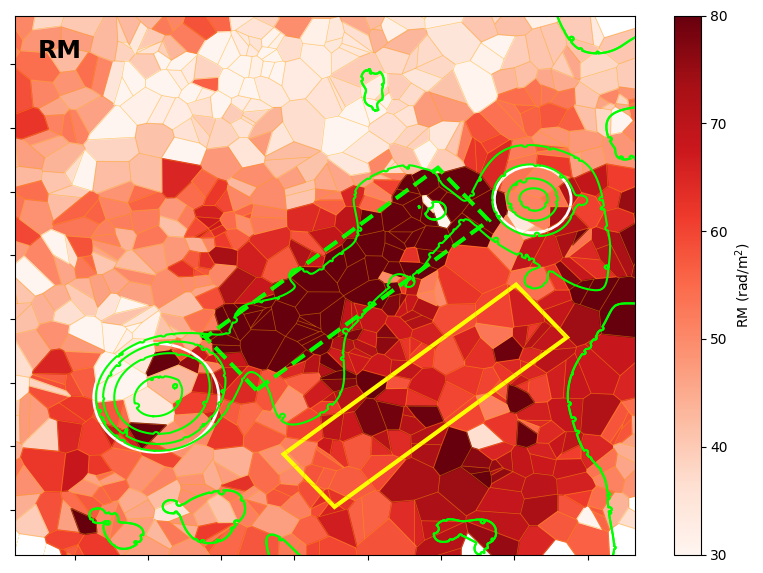}
        \includegraphics[width=0.33\linewidth]{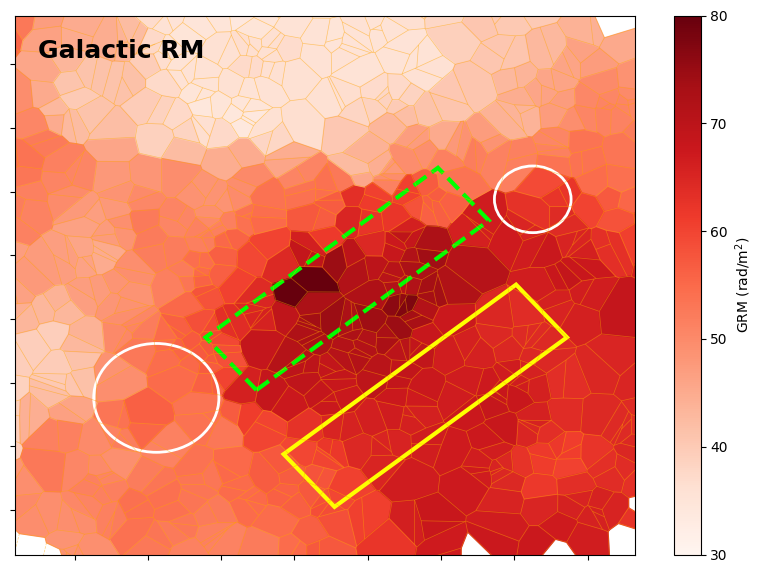}
        \includegraphics[width=0.33\linewidth]{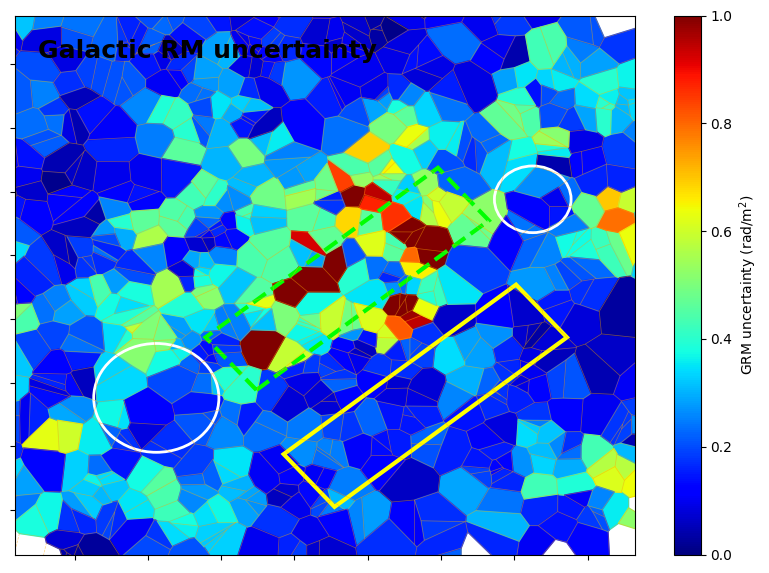}
        \caption{Voronoi visualisation of the polarised source catalogue for an injection of a distribution of RMs with <RRM>=100~rad/m$^{2}$ and $\sigma_\text{RRM}=50$~rad/m$^{2}$ in the filament region. Left: Modified RM catalogue. Centre: GRM computed with the annulus method. Right: GRM uncertainty at the position of each source. Each polygon represents a source in the catalogue. Its shape depends on the position of neighbouring sources, while its colour is described by the colour bar. The circles mark galaxy clusters, as in Fig.~\ref{fig:EMU}, and X-ray contours from \citet{Dietl24} are also shown in the top panel. The dashed green rectangle marks the filament region, while the yellow rectangle marks the control region.}
        \label{fig:voronoi_RM-GRM_testfake}
    \end{figure}

\begin{figure}[H]
        \centering
        \includegraphics[width=0.45\linewidth]{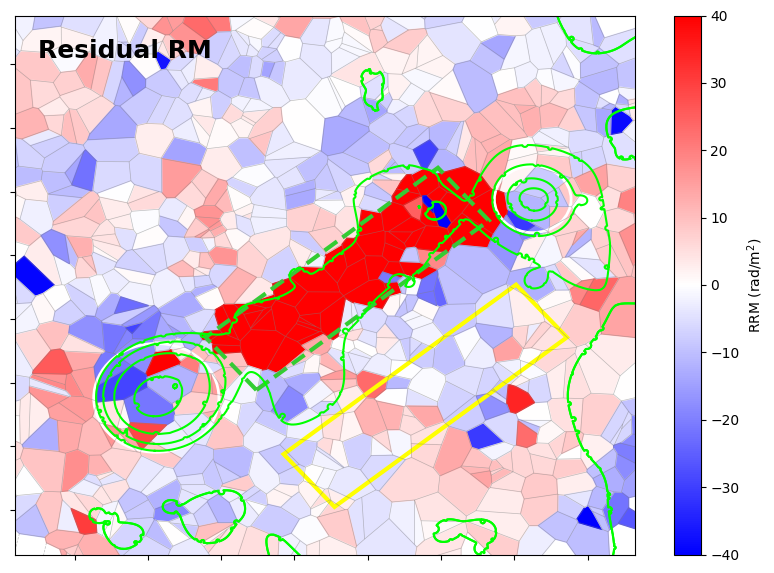}
        \includegraphics[width=0.45\linewidth]{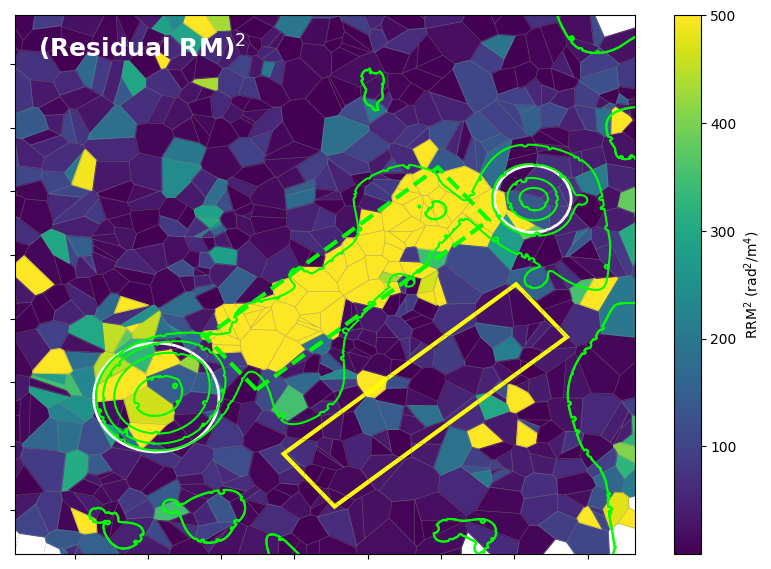}
        \caption{Voronoi visualisation of the polarised source catalogue for an injection of a distribution of RM with <RRM>=100~rad/m$^{2}$ and $\sigma_\text{RRM}=50$~rad/m$^{2}$ in the filament region. Left: RRM. Right: Its squared value. Each polygon represents a source in the catalogue. Its shape depends on the position of neighbouring sources, while its colour is described by the colour bar. The circles are the same as in Fig.~\ref{fig:EMU}, and X-ray contours from \citet{Dietl24} are also shown in the left panel. The dashed green rectangle marks the filament region, while the yellow rectangle marks the control region.}
        \label{fig:voronoi_RRM-RRM2_testfake}
    \end{figure}

In this Appendix, we perform a test to verify our method for estimating the GRM. Starting from the POSSUM RM catalogue obtained in Sect.~\ref{sec:qualitycuts}, we inject a mock distribution of RMs with $\langle \mathrm{RRM} \rangle = 100~\mathrm{rad\,m^{-2}}$ and $\sigma_\mathrm{RRM} = 50~\mathrm{rad\,m^{-2}}$ in the filament region (see Fig.~\ref{fig:voronoi_RM-GRM_testfake}, left panel). We then repeat the procedure described in Sect.~\ref{sec:GRM}, producing GRM and GRM uncertainty maps (Fig.~\ref{fig:voronoi_RM-GRM_testfake}, central and right panels), and subsequently isolate the residual extragalactic contribution (see Fig.~\ref{fig:voronoi_RRM-RRM2_testfake}). Finally, following Sect.~\ref{sec:extgalRM}, we compute the RRM statistics in both the control and filament regions. The results are reported in Table~\ref{tab:RM_testfake}.

We find that the recovered $\sigma_\mathrm{MAD,RRM}$ and $\widetilde{\mathrm{RRM}}$ are consistent with the injected values within the uncertainties. This test confirms that our method for estimating the GRM directly from POSSUM data is robust, and that an extragalactic RM signal on the scale of the filament is not significantly affected by our GRM subtraction procedure.

\begin{table}[H]
    \centering
    \caption{RRM statistics in the considered regions.}
    \small
    \begin{tabular}{cccc}
        Region & $N_{\rm sources}$ & $\sigma_\text{MAD,RRM}$ & $\widetilde{\text{RRM}}$ \\
               &             & rad/m$^{2}$ & rad/m$^{2}$  \\
               \hline
         Filament & 54 & 44.8$\pm$7.4 & 106.0$\pm$8.4 \\
         Control & 46 & 7.1$\pm$1.2 & -0.6$\pm$1.0 \\
    \end{tabular}
    \tablefoot{The values here were obtained for an injection of a mock distribution of polarised sources with <RRM>=100~rad/m$^{2}$ and $\sigma_\text{RRM}=50$~rad/m$^{2}$ in the filament region. Column 1: Region name. Column 2: Number of polarised sources in the region. Column 3: Robust estimate of the RRM standard deviation (1.48$\times$MAD). Column 4: Median RRM. Uncertainties were estimated via bootstrapping.}
    \label{tab:RM_testfake}
\end{table}

\end{appendix}

\end{document}